\documentclass[11pt]{article}
\usepackage{amsmath}
\usepackage{amsfonts}
\usepackage{amssymb}

\begin{document}
\title{\bf{Symmetries of topological gravity with torsion in the hamiltonian and lagrangian formalisms}}
\author{
{\bf {\normalsize Rabin Banerjee}}\thanks{Email: rabin@bose.res.in}\\
 {\normalsize S.~N.~Bose National Centre for Basic Sciences,}\\
 {\normalsize Block-JD, Sector III, Salt Lake, Kolkata-700098, India.}\\
\and {\bf {\normalsize Sunandan Gangopadhyay}}\thanks{E-mail: sunandan@bose.res.in, sunandan.gangopadhyay@gmail.com}\\
 {\normalsize Department of Physics and Astrophysics,}\\
 {\normalsize West Bengal State University,}\\
 {\normalsize Barasat, North 24 Paraganas, West Bengal, India.}{\footnotemark[4]}\\
\and {\bf {\normalsize Pradip Mukherjee}\thanks{E-mail: mukhpradip@gmail.com}}\\
 {\normalsize Department of Physics, Presidency College,}\\
 {\normalsize 86/1 College Street, Kolkata-700073, India.}{\footnote{Also, Visiting Associate at S.~N.~Bose National Centre for Basic Sciences, JD Block, Sector III, Salt Lake, Kolkata-700098, India.}}\\
\and {\bf {\normalsize Debraj Roy}}\thanks{E-mail: debraj@bose.res.in}\\
 {\normalsize S.~N.~Bose National Centre for Basic Sciences,}\\
 {\normalsize Block-JD, Sector III, Salt Lake, Kolkata-700098, India.}\\
}
\vspace{-0.5cm}

\date{}

\maketitle

\begin{abstract}
A systematic analysis of the symmetries of topological 3D gravity with torsion and a cosmological term, in the first order formalism, has been performed in details - both in the hamiltonian and lagrangian formalisms. This illuminates the connection between the symmetries of curved spacetime (diffeomorphisms plus local Lorentz transformations) with the Poincare gauge transformations. The Poincare gauge symmetries of the action are shown to be inequivalent to its gauge symmetries.  Finally, the complete analysis is compared with the metric formulation where the diffeomorphism symmetry is shown to be equivalent to the gauge symmetry.
\end{abstract} 

\section{Introduction}

Gravity theories in (2+1) dimensions offer an arena where one can address such subtle issues as the singularity problem or quantization, on a simpler setting \cite{Carlip:2004ba}. Interest in 3D gravity increased a lot after Witten's discovery of the equivalence of 3D gravity with a Chern-Simons gauge theory \cite{Witten:1988hc}. Inclusion of the Chern-Simons term in the Einstein-Hilbert action leads to a theory known as `topologically massive gravity' which has a massive propagating degree of freedom \cite{Deser:1981wh, Deser:1983tn, Deser:1983dr}. These theories were studied in Riemannian space time. Later, a 3D gravity theory was formulated in the Riemann-Cartan spacetime, that is with non-zero torsion \cite{Mielke:1991nn, Baekler:1992ab}. The canonical structure of the `topological 3D gravity with torsion' with a cosmological term was investigated in \cite{Blagojevic:2004hj} following Dirac's constrained hamiltonian analysis. Recently a surge of activity in various 3D gravity models has been witnessed \cite{Blagojevic:2009ek, Blagojevic:2008bn, Blagojevic:2004hj, Blagojevic:2003uc, Blagojevic:2002du, Maloney:2009ck, Carlip:2005zn, Park:2008yy, Grumiller:2008pr, Carlip:2008qh}. In this connection it may be noted that there are many subtle aspects involved in the construction of the symmetry generators and their applications, in 3D gravity models, which are not completely addressed in the literature. We will highlight some of these issues in this paper through an analysis of the topological 3D gravity model with torsion and a cosmological term.

       Gravity theories in the Riemann-Cartan space are analysed in the framework of Poincare Gauge Theory (PGT). The edifice of PGT is constructed by localising the Poincare transformations in Minkowski space. One starts with a matter theory invariant under global Poincare transformations. Naturally, this does not remain invariant when the parameters of the Poincare transformations are functions of spacetime. The PGT emerges from attempts to modify the matter theory so that it becomes invariant under the local Poincare transformations. Compensating potentials are introduced in the process, the dynamics of which is provided by invariant densities constructed out of the field strengths obtained from the usual gauge theoretic procedure \cite{Blagojevic:2002du, Utiyama:1956sy, Kibble:1961ba}. The theory has been ubiquitous in classical gravity \cite{Blagojevic:2002du} as well as in its extension to noncommutative spacetime \cite{Chamseddine:2000si, Calmet:2005qm, Mukherjee:2006nd, Banerjee:2007th}. The usefulness of the PGT stems from the fact that theories invariant under local Poincare transformations can be viewed as invariant theories in curved spacetime. The geometric interpretation of PGT, based on local Lorentz transformations (LLT) and general coordinate transformations or diffeomorphisms (diff), thus acquires a crucial significance and needs to be thoroughly understood. 
      
       The hamiltonian analysis of 3D gravity with torsion in \cite{Blagojevic:2004hj}, on the other hand, reveals that the transformations of the basic fields under the gauge generator constructed by Hamiltonian procedure differs from those obtained from PGT and agreement is only achieved by using the equations of motion \cite{Blagojevic:2008bn, Blagojevic:2004hj}. However, the fact to be noted here is, that the gauge generator itself in \cite{Blagojevic:2008bn, Blagojevic:2004hj} is constructed by an algorithm \cite{Castellani:1981us} based on the gauge symmetries that maps solutions to solutions of the equations of motion. Nonetheless, a question arises about the details of the symmetries of the theory under gauge and Poincare gauge transformations, a question that is also  relevant to the geometric interpretation mentioned above. Moreover, there are strictly hamiltonian methods of obtaining  the gauge generator without using the equations of motion \cite{Banerjee:1999hu,Banerjee:1999yc} which have been used recently \cite{Mukherjee:2007yi} in the context of second order metric gravity to establish an off-shell equivalence between the gauge and diffeomorphism parameters. The issue of the discrepancy between the Poincare gauge transformations and gauge transformations should therefore not be brushed aside. In the present paper we will try to paint the problem in its true colours.

       The problem mentioned above has indeed two aspects. The first concerns the invariance issue which rests crucially on the geometric interpretation. The PGT is a gauge theory in the Minkowski spacetime whereas gravity is formulated in curved spacetime. The principle of equivalence enables us to erect locally Lorentzian coordinates in the tangent spaces. The most general spacetime transformations comprises the general coordinate transformation or diff plus the LLT. We will establish the Poincare gauge transformations from these invariances of the curved spacetime. A thorough one to one correspondence of the geometric structures with the Poincare gauge fields will be worked out. While the geometric interpretation of PGT is well known, such elaborate analysis of the equivalence is not available in the literature. Our analysis of the geometric interpretation of PGT naturally manifests the invariance of the theory under Poincare gauge transformations. We will perform an explicit check of this invariance in our paper. As a specific example, we will consider the topological 3D gravity model which includes the Chern-Simons (CS) term, the cosmological term and torsion along with the usual Einstein-like term \cite{Blagojevic:2004hj}. 

       The second aspect is related with the construction of the gauge generator. We will follow the method of \cite{Banerjee:1999hu,Banerjee:1999yc}. Here gauge transformations are viewed as mappings of field configurations to field configurations. A strictly hamiltonian procedure is followed  and a structured algorithm is provided to obtain the independent gauge parameters. This algorithm has been applied to different reparametrization invariant models \cite{Banerjee:2004un,Gangopadhyay:2007gn,Banerjee:2006pi,Samanta:2007fk} including second order metric gravity \cite{Mukherjee:2007yi}. The gauge transformations of the basic fields obtained from this method should naturally be invariances of the action. Notwithstanding this observation, we propose an explicit check of the invariance.
     
       The 3D gravity model discussed in this paper (which is the Mielke-Baekler \cite{Mielke:1991nn, Baekler:1992ab} model along with an additional cosmological term) contains both first class and second class constraints. We provide a thorough canonical analysis of the model following Dirac's approach of constrained hamiltonian analysis. Our work supplements the  canonical analysis of \cite{Blagojevic:2004hj} in that we work out the reduced phase-space structure. This is done by eliminating the second class constraints through replacement of the Poisson brackets by Dirac brackets. The basic Dirac brackets between the fields and the conjugate momenta have been computed and used to construct the gauge generator following \cite{Banerjee:1999hu,Banerjee:1999yc}. As already mentioned this is an off-shell procedure of finding the most general gauge transformations and should be contrasted with the method of \cite{Castellani:1981us} that uses the equations of motion. The transformation of the basic fields are obtained from their Dirac brackets with the generator. We also provide an explicit check of the off-shell invariances of the action under both gauge transformations and the Poincare gauge transformations.

       At this point one interesting aspect may be noted. This is the off-shell difference between the Poincare gauge symmetries and the Hamiltonian gauge invariances.  We will show that this is indeed a peculiarity of the PGT. For example we will consider the Einstein action in the second order metric gravity formalism. The spacetime invariance of the theory is the diff transformations. In (3+1) dimensions these diff invariances have been shown to map off-shell to the hamiltonian gauge transformations \cite{Mukherjee:2007yi}. We show that the same analysis is applicable to the corresponding (2+1) dimensional theory. When we treat the same theory in the PGT framework the discrepancy between the Poincare transformations and the gauge transformations comes to the fore.

      Apart from the hamiltonian analysis, we also discuss a lagrangian treatment of both PGT symmetries and the usual gauge symmetries. Corresponding to an independent off-shell invariance of the action there exists one lagrangian gauge identity. We construct the gauge identities following from the Poincare gauge invariances and the hamiltonian gauge invariances. If the gauge identities of the first set or any combination of them can be shown to be identical with the identities of the second set, the offshell equivalence of the two different symmetries can be established. Otherwise they are inequivalent. In our case we show that the Poincare gauge invariances and the hamiltonian gauge invariances are {\it{inequivalent}} by this procedure. Since the lagrangian gauge identities are formulated in terms of the Euler derivatives of the action, they become trivial when the equations of motion are invoked. In this sense the lagrangian formulation might not be appropriate for discussing the on-shell equivalence between the two types of symmetries. It is then natural to ask whether the equivalence of the spacetime symmetries of the Einstein gravity with its gauge invariances can be demonstrated from the lagrangian point of view. We answer the question in the affirmative by showing the equivalence of the corresponding gauge/diff identities.
 
    The organisation of the paper is as follows. In the next section we give a short review of PGT and introduce the three dimensional topological gravity model with torsion and a cosmological term \cite{Blagojevic:2004hj, Blagojevic:2003uc}. In Section 3 the Poincare gauge transformations of the basic fields are established from the geometric correspondence. The invariance of our 3D gravity model, introduced in the previous section, under the Poincare gauge transformations will be examined here. Section 4 contains a thorough canonical analysis of our model  including the reduced phase-space structure. Using the results of Section 4 the generator of the general gauge transformations is constructed and the transformations of the basic fields under the gauge transformation are provided in section 5. Note that we follow an off-shell method of construction of the gauge generator \cite{Banerjee:1999hu,Banerjee:1999yc} which maps field configurations to equivalent field configurations. The off-shell invariance of the model under the gauge transformations is thus expected. We provide an explicit verification of the same in this section. Also, after being sure about the invariances from different angles, we compare the Poincare gauge transformations and gauge transformations of the basic fields. This fulfills our motivation of viewing the difference between both types of transformations in the proper perspective, a fact noticed in the literature in recent times \cite{Blagojevic:2002du, Blagojevic:2004hj, Blagojevic:2003uc} but not sufficiently highlighted. We will give some definitive arguments in section 6 to show that this is a peculiarity of the Poincare gauge theory framework. This is done by analysing the 3D Einstein action in the 2nd order formalism. Here spacetime invariances are the diffeomorphisms. We show that they are equivalent {\it{off-shell}} to the gauge transformations obtained by a hamiltonian procedure following the analysis in \cite{Mukherjee:2007yi}. We then consider the same action in the PGT framework and see that the equivalence is lost.  Our canonical analysis amply demonstrates that if the Poincare gauge transformations were equivalent to gauge transformations there would exist some off-shell mapping between them. To further elucidate the point we resort to a lagrangian framework. The distinctive features between the two sets of transformations is confirmed by constructing identities involving the Euler derivatives for both the symmetry transformations. These identities are inequivalent in the sense that no mapping exists between them. In the connection of the lagrangian analysis one wonders about the 2nd order approach. We explicitly show that in case of the 3D second order metric gravity the gauge identities corresponding to the gauge transformations can be mapped to the corresponding identities following from reparametrization, thereby reinforcing the off-shell equivalence between the two symmetries. Finally, we conclude in  Section 7. 

      Before concluding the introductory section, a summary of  our conventions regarding the indices will be appropriate. Latin indices refer to the local Lorentz frame and the Greek indices refer to the coordinate frame. The first letters of both alphabets $(a,b,c,\ldots)$ and $(\alpha,\beta,\gamma,\ldots)$ run over 1,2 while the middle alphabet letters $(i,j,k,\ldots)$ and $(\mu,\nu,\lambda,\ldots)$ run over 0,1,2. The totally antisymmetric tensor densities $\varepsilon^{ijk}$ and  $\varepsilon^{\mu\nu\rho}$ are both normalized so that $\varepsilon^{012}=1$. The signature of spacetime is $\eta=(+,-,-)$.
         
\section{Topological 3D gravity with torsion}

The topological 3D gravity theory with torsion is formulated in (2+1) dimensional Riemann-Cartan spacetime in the framework of the Poincare Gauge Theory (PGT). The starting point of the PGT is a matter theory invariant under the global Poincare transformations in the Minkowski space:
\begin{equation}
x^\mu \rightarrow x^\mu + \xi^\mu\label{GPT}
\end{equation}
where $\xi^\mu = \theta^\mu_{\ \nu}x^\nu + \epsilon^\mu$, with both $\theta^{\mu\nu}$ and $\epsilon^\mu$ being constants. At each spacetime point a local basis ${\bf{e}}_i$ is considered which are related to the coordinate basis ${\bf{e}}_\mu$ by $${\bf{e}}_i = \delta^\mu_i{\bf{e}}_\mu.$$ When we make the global Poincare symmetry (\ref{GPT}) local, the transformation parameters $\theta^{ij}$ and $\epsilon^\mu$ become functions of the spacetime coordinates. In this case one can take $\xi^\mu = \theta^\mu_{\ \,\nu} x^\nu +  \epsilon^\mu$ and $\theta^{ij}$ as the independent parameters. To ensure the invariance of the matter action under the {\it{local}} Poincare transformations, compensating fields $b^i_{\ \mu}$ and $\omega^{ij}_{\ \ \mu}$ are required to be introduced \cite{Blagojevic:2002du}. The covariant derivative $\nabla_k$ is constructed using these fields as $$\nabla_k = b_k^{\ \mu}\nabla_\mu$$ where $\nabla_\mu = \partial_\mu + \frac{1}{2}\omega^{ij}_{\ \ \mu}\Sigma_{ij}$ will be called the $\theta$-covariant derivative. The matrix $\Sigma_{ij}$ is the Lorentz spin matrix and $b_k^{\ \mu}$ is the inverse to $b^k_{\ \mu}$. The covariant derivative of the matter field $\phi$ is required to transform under local Poincare transformations just as the ordinary derivative $\partial_{k}\phi$ transforms under global Poincare transformations. The matter lagrangian density ${\cal{L}} = {\cal{L}}(\phi,\partial_{k}\phi)$ which was invariant under global Poincare transformations is converted to an invariant density ${\cal{\tilde{L}}}$ by replacing the ordinary derivative $\partial_{k}$ by the covariant derivative $\nabla_k $ i.e. $${\cal{\tilde{L}}} = {\cal{\tilde{L}}}(\phi,\nabla_{k}\phi).$$ An invariant action is constructed from this density as $$I = \int b{\cal{\tilde{L}}}(\phi,\nabla_{k}\phi)$$ where $b = {\rm det}\,b^i_{\ \mu}.$ This action is invariant under the following transformations of the basic fields $b^i_{\ \mu}$ and $\omega^{ij}_{\ \ \mu}$:
\begin{eqnarray}
\delta b^i_{\ \mu} &=& \theta^i_{\ k} b^k_{\ \mu} - \partial_\mu\xi^\rho b^i_{\ \rho} - \xi^{\rho}\partial_{\rho} b^i_{\ \mu}\nonumber\\
\delta \omega^{ij}_{\ \ \mu} &=& \theta^i_{\ k} \omega^{kj}_{\ \ \mu} + \theta^j_{\ k} \omega^{ik}_{\ \ \mu} - \partial_\mu\theta^{ij} - \partial_\mu\xi^\rho \omega^{ij}_{\ \ \rho} - \xi^{\rho}\partial_{\rho}\omega^{ij}_{\ \ \mu}\label{fieldtrans}
\end{eqnarray}
These transformations (\ref{fieldtrans}) comprise the Poincare gauge transformations. Their structure suggests a geometric interpretation. The basic fields $b^i_{\ \mu}$ and $\omega^{ij}_{\ \ \mu}$ mimic the triad and the spin connection in curved spacetime. The most general invariance group in curved spacetime consists of the LLT plus diff. Observe in (\ref{fieldtrans}), that the Latin indices transform as under LLT with parameters $\theta^{ij}$, and the Greek indices transform as under diff with parameters $\xi^\mu$. This suggests a correspondence between the Poincare gauge transformations with the geometric transformations of the curved spacetime. However, that this correspondence is an equivalence, is by no means obvious. In the next section, we will precisely establish this equivalence by obtaining the Poincare gauge transformations starting from the geometric (LLT + diff) invariances. Such an explicit map between the two sets of transformations is essential because the geometric interpretation of PGT is a crucial step. It enables us to cast gravity in the framework of PGT. 
 
With the geometric correspondence lurking behind, we come back to the construction of PGT. Corresponding to the basic fields $b^i_{\ \mu}$ and $\omega^{ij}_{\ \ \mu}$ the Lorentz field strength $R^{ij}_{\ \ \mu\nu}$ and the translation field strength $T_{i\mu\nu}$ are obtained following the usual procedure in gauge theory. The commutator of two $\theta$-covariant derivatives gives $R^{ij}_{\ \ \mu\nu}$
$$ \left[\nabla_\mu,\nabla_\nu\right]\phi = \frac{1}{2}R^{ij}_{\ \ \mu\nu}\Sigma_{ij}\phi$$
whereas the commutator of two $\nabla_k$ derivatives furnish the additional fields $T_{i\mu\nu}$ as
$$ \left[\nabla_k,\nabla_l\right]\phi = \frac{1}{2} ~b_k^{\ \mu} b_l^{\ \nu} ~R^{ij}_{\ \ \mu\nu}\Sigma_{ij}\phi - b_k^{\ \mu}b_l^{\ \nu}~T^i_{\ \mu\nu}\nabla_i\phi$$
These defining equations give the following expressions for the field-strengths
\begin{eqnarray}
\label{pgt}
T^i_{\ \mu\nu} &=& \partial_\mu b^i_{\ \nu} + \omega^{i}_{\ \, k\mu} b^k_{\ \nu} - \partial_\nu b^i_{\ \mu} - \omega^{i}_{\ \,k\nu} b^k_{\ \mu}\nonumber\\
R^{ij}_{\ \ \mu\nu} &=& \partial_\mu \omega^{ij}_{\ \ \nu} - \partial_\nu \omega^{ij}_{\ \ \mu} + \omega^i_{\ k\mu}\omega^{kj}_{\ \ \nu} - \omega^i_{\ k\nu}\omega^{kj}_{\ \ \mu}.
\end{eqnarray}
So far the theory is in the Minkowski space and has been developed as a gauge theory. From the point of view of geometric interpretation, the Lorentz field strength $R^{ij}_{\ \ \mu\nu}$ and the translation field strength $T_{i\mu\nu}$, correspond to the Riemann tensor and the torsion. Using these basic structures, gravity can be formulated in the framework of PGT. In three dimensions, the following maps are used to simplify the analysis
\begin{eqnarray}
\label{3fields}
\theta_i &=& \frac{1}{2}\varepsilon_{ijk}\,\theta^{jk}\nonumber\\
\omega^i_{\ \mu} &=& -\frac{1}{2}\varepsilon^i_{\ jk}\,\omega^{jk}_{\ \ \mu}\nonumber\\
R_{i\mu\nu} &=& -\frac{1}{2}\varepsilon_{ijk}\,R^{jk}_{\ \ \mu\nu}
\label{simplifications}
\end{eqnarray}
Using this map in the expressions (\ref{pgt}) for $R^{ij}_{\ \ \mu\nu}$ and $T_{i\mu\nu}$ we can write
\begin{eqnarray}
\label{Tdefn}
T_{i\nu\rho}&=&\partial_\nu b_{i\rho} + \varepsilon_{ijk} \omega^j_{\ \nu} b^k_{\ \rho} - \partial_\rho b_{i\nu} - \varepsilon_{ijk} \omega^j_{\ \rho} b^k_{\ \nu }\\
\label{Rdefn}
R_{i\nu\rho}&=&\partial_\nu \omega_{i\rho}-\partial_\rho \omega_{i\nu} + \varepsilon_{ijk} \omega^j_{\ \nu} \omega^k_{\ \rho}.
\end{eqnarray}
These expressions are used to construct our action for the topological 3D gravity model with torsion and a cosmological term, without higher order derivatives \cite{Mielke:1991nn, Baekler:1992ab, Blagojevic:2003uc}
\begin{eqnarray}
\label{action}
I=\int ~d^3x~ \varepsilon^{\mu\nu\rho}\left[ab^i_{\ \mu} R_{i\nu\rho}-\frac{\Lambda}{3} \varepsilon_{ijk}b^i_{\ \mu} b^j_{\ \nu} b^k_{\ \rho} + \alpha_3 \left(\omega^i_{\ \mu} \partial_\nu\omega_{i\rho} + \frac{1}{3} \varepsilon_{ijk}\omega^i_{\ \mu} \omega^j_{\ \nu} \omega^k_{\ \rho} \right) + \frac{\alpha_4}{2} b^i_{\ \mu} T_{i\nu\rho} \right].
\end{eqnarray}
Here $a$, $\Lambda$, $\alpha_3$ and $\alpha_4$ are arbitrary parameters. The first term proportional to $a$ is the Einstein-Hilbert action written in three dimensions using the identity
\begin{equation}
\label{theidentity}
bR = -\varepsilon^{\mu\nu\rho}\,b^i_{\ \mu} R_{i\nu\rho}
\end{equation}
where $b = \text{det}\,b^i_{\ \mu}$ and $R = b_i^{\ \mu} b_j^{\ \nu} R^{ij}_{\ \ \mu\nu}$. The second term is the cosmological constant part, the third one is the Chern-Simons action while the fourth includes torsion. These terms can be manipulated with the help of the adjustable parameters $a$, $\Lambda$, $\alpha_3$ and $\alpha_4$. 
 
The variations of the action (\ref{action}) w.r.t the triad $b^i_{\ \mu}$ and the spin connection $\omega^i_{\ \mu}$ are given by
\begin{equation}
\frac{\delta I}{\delta b^i_{\ \mu}}= -\varepsilon^{\mu\nu\rho}\left[aR_{i\nu\rho} +\alpha_4T_{i\nu\rho} - \Lambda\varepsilon_{ijk}b^{i\nu}b^{k\rho}\right]
\end{equation}
\begin{equation}
\frac{\delta I}{\delta \omega^i_{\ \mu}}= -\varepsilon^{\mu\nu\rho}\left[\alpha_3R_{i\nu\rho} + aT_{i\nu\rho} - \Lambda\varepsilon_{ijk}b^{i\nu}b^{k\rho}\right]
\end{equation}
These Euler derivatives will serve a twofold purpose. First, their vanishing yields the equations of motion in the usual way. Secondly, they will appear in the lagrangian gauge identities to be discussed in section 6. The equations of motion following from the action can be simplified in the sector\footnote{Later in section 4 we will find that this condition is essential in computing the Dirac brackets.\label{fn:neq 0 condition}} $$\alpha_3\alpha_4 - a^2 \neq 0$$ as
\begin{eqnarray}
\label{eqnmot}
T^i_{\ \mu\rho} - p\,\epsilon^{ijk}\,b_{j \mu}b_{k\rho} &=& 0\nonumber\\
R^i_{\ \mu\rho} - q\,\epsilon^{ijk}\,b_{j \mu}b_{k\rho} &=& 0.
\end{eqnarray}
where $p=\dfrac{\alpha_3\Lambda + \alpha_4 a}{\alpha_3\alpha_4 - a^2}$ and $q=-\dfrac{\alpha_4^{\ 2} +a\Lambda}{\alpha_3\alpha_4 - a^2}$. 

\section{On the geometric interpretation of the Poincare gauge transformations}

   In this section we will establish the Poincare gauge transformations from the invariances of the curved spacetime. In curved spacetime erect local basis ${\bf{b_i (x^\mu)}}$. The coordinate basis ${\bf{b_\mu }}$ is derived from the tangents to the coordinate lines. The connection with the local basis is established by the triad and the parallel transport is defined in the local basis by the spin connection. Geometric structures such as the metric, affine connection etc. can be established in terms of the triad and spin connection by certain very general requirements (invariance of the interval, vielbein postulate etc). The metric is defined as 
\begin{equation}
\label{metric}
g_{\mu\nu} = \eta_{ij}b^i_{\ \mu} b^j_{\ \nu}
\end{equation}
The geometric invariances in the curved spacetime comprise the diffeomorphism (diff) and local Lorentz transformations (LLT). Under diff, the variation of $g_{\mu\nu}$ is given by
\begin{equation}
\label{metricvar}
\delta g_{\mu\nu} =  -\partial_\mu\xi^\rho g_{\rho\nu}- \partial_\nu\xi^\rho g_{\mu\rho} - \xi^\rho\partial_\rho g_{\mu\nu}
\end{equation}
Using (\ref{metric}), the lhs of (\ref{metricvar}) can be expressed in terms of the variation $\delta b^i_{\ \mu}$ as 
\begin{equation}
\label{metricvarlhs}
\delta g_{\mu\nu} = \delta\eta_{ij}b^i_{\ \mu}b^j_{\ \nu}+\eta_{ij}\,\delta b^i_{\ \mu}\,b^j_{\ \nu} + \eta_{ij}b^i_{\ \mu}\,\delta b^j_{\ \nu}
\end{equation}
where $\delta\eta_{ij}$ is the variation of $\eta_{ij}$ under LLT, given by
\begin{equation}
\label{metricidentity}
\delta\eta_{ij} = \theta_i^{\ k}\eta_{kj} + \theta_j^{\ k}\eta_{ik} = 0.
\end{equation}
Equating (\ref{metricvar}) with (\ref{metricvarlhs}) we get
\begin{eqnarray}
\label{metricvarstep1}
\eta_{ij} \,\delta b^i_{\ \mu}\,b^j_{\ \nu} + \eta_{ij}b^i_{\ \mu}\,\delta b^j_{\ \nu} \!\!&+&\!\! \theta_i^{\ k}\eta_{kj}b^i_{\ \mu} b^j_{\ \nu} + \theta_j^{\ k}\eta_{ik}b^i_{\ \mu} b^j_{\ \nu} \\ \nonumber
\!&=&\! -\partial_\nu\xi^\rho \eta_{ij}b^i_{\ \mu} b^j_{\ \rho}-\partial_\mu\xi^\rho \eta_{ij}b^i_{\ \rho} b^j_{\ \nu} - \xi^\rho\partial_\rho \left(b^i_{\ \mu}b^j_{\ \nu}\right) \eta_{ij}
\end{eqnarray}
Simplification yields
\begin{eqnarray}
\label{metricvarstep2}
b_{i\nu}\left[\delta b^i_{\ \mu} + \partial_\mu\xi^\rho b^i_{\ \rho} + \xi^\rho\partial_\rho b^i_{\ \mu} + \theta_k^{\ i}b^k_{\ \mu}\right] + b_{j\mu}\left[\delta b^j_{\ \nu} + \partial_\nu\xi^\rho b^j_{\ \rho} + \xi^\rho\partial_\rho b^j_{\ \nu} + \theta_k^{\ j}b^k_{\ \nu}\right]= 0
\end{eqnarray}
The last equation leads to
\begin{eqnarray}
\label{metricvarstep3}
\delta b^i_{\ \mu} = \theta^i_{\ k} b^k_{\ \mu} - \partial_\mu\xi^\rho b^j_{\ \rho} - \xi^\rho\partial_\rho b^i_{\ \mu} 
\end{eqnarray}
This is identical with the first set of (\ref{fieldtrans}). Note that the variation $\delta\eta_{ij}$ of the constant tensor $\eta_{ij}$ under LLT, given by (\ref{metricidentity}), reproduces the expected vanishing result. Nevertheless it has to be included and split in (\ref{metricvarstep2}) in order to get agreement with the corresponding Poincare gauge transformation (\ref{fieldtrans}). Otherwise, the $\theta$-contribution in (\ref{metricvarstep3}) will be lacking. This observation further elucidates the connection between geometric transformations and those of PGT.

In order to reproduce the second set of (\ref{fieldtrans}), consider the transformation of the affine connection $\Gamma^\mu_{\nu\lambda}$. Using the vielbein postulate we can write
\begin{equation}
\label{affcon}
\Gamma^\mu_{\nu\lambda} = b_i^{\ \mu}\partial_\lambda b^i_{\ \nu} + \omega^i_{\ j\lambda}b_i^{\ \mu} b^j_{\ \nu}
\end{equation}
It transforms under diff as 
\begin{equation}
\delta\Gamma^\mu_{\nu\lambda} = -\partial_\nu\xi^\rho~\Gamma^\mu_{\rho\lambda} - \partial_\lambda\xi^\rho~\Gamma^\mu_{\nu\rho} + \partial_\rho\xi^\mu~\Gamma^\rho_{\nu\lambda} -\partial_\nu\partial_\lambda\xi^\mu - \xi^\rho~\partial_\rho\Gamma^\mu_{\nu\lambda}
\label{conalj1}
\end{equation}
On the other hand, from (\ref{affcon}), we obtain
\begin{eqnarray}
\delta \Gamma^\mu_{\nu\lambda} = \delta b_i^{\ \mu}\,\partial_\lambda b^i_{\ \nu} + b_i^{\ \mu}\,\partial_\lambda \delta b^i_{\ \nu} + \delta\omega^i_{\ j\lambda}\,b_i^{\ \mu} b^j_{\ \nu} + \omega^i_{\ j\lambda}\,\delta b_i^{\ \mu}\,b^j_{\ \nu} + \omega^i_{\ j\lambda}b_i^{\ \mu}\,\delta b^j_{\ \nu}
\label{varGamma19}
\end{eqnarray}
Equating (\ref{conalj1}) with (\ref{varGamma19}) and using (\ref{metricvarstep3}), one finds after a long algebra,
\begin{eqnarray}
\delta \omega^{ij}_{\ \ \mu} &=& \theta^i_{\ k} \omega^{kj}_{\ \ \mu} + \theta^j_{\ k} \omega^{ik}_{\ \ \mu} - \partial_\mu\theta^{ij} - \partial_\mu\xi^\rho \omega^{ij}_{\ \ \rho} - \xi^{\rho}\partial_{\rho}\omega^{ij}_{\ \ \mu}
\label{conalj2}
\end{eqnarray}
which is equivalent to the second set of (\ref{fieldtrans}). We thus find that if we identify the triad $b^i_{\ \mu}$ and the spin connection $\omega^{ij}_{\ \ \lambda}$ with the `gauge potentials' $b^i_{\ \mu}$ and  $\omega^{ij}_{\ \ \lambda}$ of the PGT, then spacetime symmetry transformations (namely, the LLT and diff) generate the same transformations as the Poincare gauge transformations. 

The above correspondences show the connection of Poincare gauge symmetry with LLT plus diff invariances in curved spacetime. This connection may further be pursued at the level of the field strengths. From the point of view of PGT the transformations of the Lorentz field strength $R^{ij}_{\ \ \mu\nu}$ and the translation field strength $T^i_{\ \mu\nu}$ can easily be obtained by direct substitution of (\ref{fieldtrans}) in (\ref{pgt}),
\begin{eqnarray}
\label{lorentztrans}
\delta R^{ij}_{\ \ \mu\nu} = \theta^i_{\ k}\,R^{kj}_{\ \ \mu\nu} +\theta ^j_{\ k}\,R^{ik}_{\ \ \mu\nu} -\partial_{\mu}\xi^{\rho}\,R^{ij}_{\ \ \rho\nu} - \partial_{\nu}\xi^{\rho}\,R^{ij}_{\ \ \mu\rho} -\xi^{\rho}\,\partial_{\rho}R^{ij}_{\ \ \mu\nu}
\end{eqnarray}
and
\begin{eqnarray}
\label{transtrans}
\delta T^{i}_{\ \mu\nu} = \theta ^i_{\ k}\,T^{k}_{\ \ \mu\nu} -\partial_{\mu}\xi^{\rho}\,T^{i}_{\ \rho\nu} - \partial_{\nu}\xi^{\rho}\,T^{i}_{\ \mu\rho} - \xi^{\rho}\,\partial_{\rho}T^{i}_{\ \mu\nu}
\end{eqnarray}
As noted earlier in the context of the transformations (\ref{fieldtrans}), the Latin indices transform as under LLT with parameters $\theta^{ij}$ and the Greek indices transform as under diff with parameters $\xi^\mu$. We get the expected transformations under LLT and diff. This agreement lends further support for viewing PGT as gravity theory on curved spacetime.

Before we pass on to the invariance of the model it will be convenient to write (\ref{fieldtrans}) specific to 3D. This is achieved by using the map (\ref{3fields}) in (\ref{fieldtrans}). The resultant transformation relations are
\begin{eqnarray}
\label{fieldtrans3D}
\delta b^i_{\ \mu} &=& \theta^i_{\ k} b^k_{\ \mu} - \partial_\mu\xi^\rho \, b^i_{\ \rho} - \xi^{\rho}\,\partial_{\rho}b^i_{\ \mu} \nonumber\\
\delta \omega^{i}_{\ \mu} &=& -\left(\partial_\mu\theta^i + \varepsilon^i_{\ jk}\omega^j_{\ \mu}\theta^k \right) - \partial_\mu\xi^\rho \, \omega^i_{\ \rho} - \xi^{\rho}\,\partial_{\rho}\omega^i_{\ \mu}
\end{eqnarray}

   From the analysis of the geometric correspondence it is natural to expect the action (\ref{action}) to be invariant under the Poincare gauge transformations (\ref{fieldtrans3D}). A straightforward calculation leads to the following variation $\delta I$ of the action:
\begin{equation}
\delta I =\delta I^{(1)} + \delta I^{(2)}\label{actiontrans}
\end{equation}
where
\begin{eqnarray}
\label{invariance1}
\delta I^{(1)} = \int d^3x \!\!\!\!&\!&\!\! \partial_\lambda \left[\xi^\lambda\,\varepsilon^{\mu\nu\rho} \left\lbrace - ab^i_{\ \mu} R_{i\nu\rho} -\alpha_3\big(\omega^i_{\ \mu}\partial_\nu\omega_{i\rho} + \frac{1}{3}\varepsilon_{ijk} \omega^i_{\ \mu} \omega^j_{\ \nu} \omega^k_{\ \rho}\big) + \Lambda\varepsilon_{ijk}b^i_{\ \mu}b^j_{\ \nu}b^k_{\ \rho} \right.\right. \\ \nonumber
&\,& \left.\left.\qquad\qquad\quad - \dfrac{\alpha_4}{2} b^i_{\ \mu}T_{i\nu\rho}\right\rbrace + \theta^i\varepsilon^{\lambda\nu\rho} \,\partial_\nu \omega_{i\rho} \right]
\end{eqnarray}
and
\begin{eqnarray}
\delta I^{(2)} &=& \int d^3x ~\varepsilon^{\mu\nu\rho}\left[a\,\left\lbrace\partial_\mu\xi^\lambda\, b^i_{\ \lambda}R_{i\nu\rho} +\partial_\nu\xi^\lambda\, b^i_{\ \mu}R_{i\lambda\rho} +\partial_\rho\xi^\lambda\, b^i_{\ \mu}R_{i\nu\lambda} -\partial_\lambda\xi^\lambda\, b^i_{\ \mu}R_{i\nu\rho}\right\rbrace\right. \nonumber\\
&\,&\qquad-\, \frac{\Lambda}{3} \varepsilon_{ijk} \left\lbrace -3\,\partial_\mu\xi^\lambda\, b^i_{\ \lambda}b^j_{\ \mu}b^k_{\ \rho} + \partial_\lambda\xi^\lambda\, b^i_{\ \mu}b^j_{\ \nu}b^k_{\ \rho} \right\rbrace + \alpha_3\left\lbrace \partial_\mu\xi^\mu (\omega^i_{\ \lambda}\partial_\nu\omega_{i\rho} +\frac{1}{3}\varepsilon_{ijk}\omega^i_{\ \lambda}\omega^j_{\ \nu}\omega^k_{\ \rho})\right. \nonumber\\
&\ &\qquad\qquad +\left. \partial_\nu\xi^\lambda\omega^i_{\ \mu} \partial_\lambda\omega_{i\rho} + \partial_\rho\xi^\lambda\omega^i_{\ \mu}\partial_\nu\omega_{i\lambda} - \partial_\lambda\xi^\lambda ( \omega^i_{\ \mu}\partial_\nu\omega_{i\rho} + \frac{1}{3}\varepsilon_{ijk}\omega^i_{\ \mu}\omega^j_{\ \nu}\omega^k_{\ \rho})\right\rbrace \nonumber\\
&\,&\qquad+\, \left.\frac{\alpha_4}{2}\left\lbrace\partial_\mu\xi^\lambda b^i_{\ \lambda}T_{i\nu\rho} + \partial_\nu\xi^\lambda b^i_{\ \mu}T_{i\lambda\rho} + \partial_\rho\xi^\lambda b^i_{\ \mu}T_{i\nu\lambda} - \partial_\lambda\xi^\lambda b^i_{\ \mu}T_{i\nu\rho}\right\rbrace\right]
\label{invariance2}
\end{eqnarray}
The piece $\delta I^{(1)}$ is a total boundary term but $\delta I^{(2)}$ is not so. The latter actually vanishes. To see this in a compact manner we use the identity following from the transformation of the tensor density $\varepsilon^{\mu\nu\rho}$. We have
\begin{equation}
\delta\varepsilon^{\mu\nu\rho} = \partial_\lambda\xi^\mu\varepsilon^{\lambda\nu\rho} +\partial_\lambda\xi^\nu\varepsilon^{\mu\lambda\nu}+\partial_\lambda\xi^\rho\varepsilon^{\mu\nu\lambda}-\partial_\lambda\xi^\lambda\varepsilon^{\mu\nu\rho} = 0
\label{epstrans}
\end{equation}
since $\varepsilon^{\mu\nu\rho}$ is a constant tensor density. Now the rhs of (\ref{invariance2}) simplifies as,
\begin{equation}
\delta I^{(2)} = \int d^3x\, \delta\varepsilon^{\mu\nu\rho} \left[ab^i_{\ \mu} R_{i\nu\rho}-\frac{\Lambda}{3} \varepsilon_{ijk}b^i_{\ \mu} b^j_{\ \nu} b^k_{\ \rho} + \alpha_3 \left\lbrace\omega^i_{\ \mu} \partial\omega^i_{\ \rho} + \frac{1}{3} \varepsilon_{ijk}\omega^i_{\ \mu} \omega^j_{\ \nu} \omega^k_{\ \rho} \right\rbrace + \frac{\alpha_4}{2} b^i_{\ \mu} T_{i\nu\rho} \right]
\end{equation}
and hence, $\delta I^{(2)}$ vanishes on account of (\ref{epstrans}). The invariance of the theory (\ref{action}) under Poincare gauge transformations (\ref{fieldtrans3D}) is thus explicitly verified.
 
\section{Canonical analysis of the model}

In considering the 2+1 dimensional model (\ref{action}) with the Chern-Simons term along with the torsion, cosmological and usual Einstein-Hilbert terms, we actually get a mixed system with both first-class and second-class constraints. This calls for a more general analysis than what is done for pure gauge systems with only first-class constraints. Such mixed systems can be  dealt by different approaches.
\begin{itemize}
\item {\bf\emph{Using Poisson brackets:}} In this method, the entire algebra is computed using Poisson brackets. Second class constraints are taken care by introducing Lagrange multipliers which enforce these constraints. The multipliers can be fixed from the time conservation of the constraints.
\item {\bf\emph{Using Dirac brackets:}} The second-class constraints can be strongly eliminated by using Dirac brackets,  and we can deal with an effectively pure system having only first-class constraints. All Poisson brackets will have to be replaced by corresponding Dirac brackets.
\end{itemize}
Here in this paper, we adopt the method of using Dirac brackets due to two reasons. First, the analysis of this model  through Dirac brackets is new and provides an interesting alternative to other studies \cite{Blagojevic:2004hj} on this model. Secondly, the systematic method of computing a generator from a structured algorithm \cite{Banerjee:1999hu, Banerjee:1999yc}, which is adopted here, is technically simple for pure systems. Thus it is desirable, though not essential, to first convert our mixed system into a pure gauge system.

The action (\ref{action}) is written in terms of the triads $b^i_{\ \mu}(x)$ and spin connections $\omega^i_{\ \mu}(x)$, which are the basic fields in this theory. The corresponding momenta $\pi_i^{\ \mu}(x)$ and $\Pi_i^{\ \mu}$, defined as $\frac{\partial \mathcal{L}}{\partial(\partial_0 b^i_{\ \mu})}$ and $\frac{\partial \mathcal{L}}{\partial(\partial_0 \omega^i_{\ \mu})}$ respectively, are found to be,
\begin{eqnarray}
\label{primary const}
\begin{split}
\phi^{\ 0}_i &= \pi^{\ 0}_i \approx 0\\
\phi^{\ \alpha}_i &=  \pi_i^{\ \alpha}-\alpha_4 \varepsilon^{0\alpha\beta}b_{i\beta}\approx 0\\
\Phi^{\ 0}_i &= \Pi^{\ 0}_i \approx 0\\
\Phi^{\ \alpha}_i &= \Pi_i^{\ \alpha} - \varepsilon^{0\alpha\beta}\left(2ab_{i\beta} + \alpha_3 \omega_{i\beta} \right)\approx 0.\\
\end{split}
\end{eqnarray}
We now see that \emph{all} the momenta lead to constraints. These are the primary constraints of the theory, defined by $\phi^{\ 0}_i$, $\phi^{\ \alpha}_i$, $\Phi^{\ 0}_i$ and $\Phi^{\ \alpha}_i$. The symbol $\approx$ stands for weak equality in the sense of Dirac \cite{Dirac:Lectures} implying that the constraints can be set equal to zero only after computing all relevant brackets.

The canonical hamiltonian density $\mathcal{H}_C$ can now be written down through a Legendre transformation $\mathcal{H}_C=\pi^{\ \mu}_i \,\partial_0 b^i_{\ \mu} + \Pi^{\ \mu}_i \,\partial_0 \omega^i_{\ \mu} - \mathcal{L}$,
\begin{eqnarray}
\label{canon Hamilt}
\begin{split}
\mathcal{H}_C &= b^i_{\ 0}\mathcal{H}_i + \omega^i_{\ 0} \mathcal{K}_i + \partial_\alpha D^\alpha\\
\mathcal{H}_i &= -\varepsilon^{0\alpha\beta}\left(a\,R_{i\alpha\beta}+\alpha_4 \,T_{i\alpha\beta}-\Lambda \,\varepsilon_{ijk} \,b^j_{\ \alpha} b^k_{\ \beta} \right)\\
\mathcal{K}_i &= -\varepsilon^{0\alpha\beta}\left(a\,T_{i\alpha\beta}+\alpha_3 \,R_{i\alpha\beta}+\alpha_4 \,\varepsilon_{ijk} \,b^j_{\ \alpha} b^k_{\ \beta} \right)\\
D^\alpha &= \varepsilon^{0\alpha\beta} \left[\omega^i_{\ 0}\left(2a\,b^i_{\ \beta} + \alpha_3\,\omega_{i\beta}\right)+ \alpha_4 \,b^i_{\ 0} b_{i\beta}\right].\\
\end{split}
\end{eqnarray}
From here, we can write down a total hamiltonian density $\mathcal{H}_T$
\begin{eqnarray}
\label{total Hamilt}
\mathcal{H}_T &=& \mathcal{H}_C + \lambda^{(3) i}_{\ \ \ \ \mu} \,\phi^{\ \mu}_i + \lambda^{(4) i}_{\ \ \ \ \mu} \,\Phi^{\ \mu}_i\nonumber\\
&=& b^i_{\ 0} \,\bar{\mathcal{H}_i} + \omega^i_{\ 0} \,\bar{\mathcal{K}_i} + \lambda^{(3) i}_{\ \ \ \ 0} \,\pi_i^{\ 0} + \lambda^{(4) i}_{\ \ \ \ 0} \,\Pi_i^{\ 0} + \partial_\alpha \bar{D^\alpha},
\end{eqnarray}
where $\lambda^{(3)}$ and $\lambda^{(4)}$ are undetermined multipliers and the quantity $\bar{D^\alpha}$ is defined as
\begin{eqnarray}
\bar{D^\alpha} = D^\alpha + b^i_{\ 0} \phi_i^{\ \alpha} + \omega^i_{\ 0} \phi_i^{\ \alpha}.
\end{eqnarray}
On using the the time conservation conditions of the primary constraints, we find  two secondary constraints,
\begin{eqnarray}
\label{second const}
\begin{split}
\bar{\mathcal{H}_i} &:=\mathcal{H}_i - \nabla_\beta \phi_i^{\ \beta} + \varepsilon_{ijk}\,b^j_{\ \beta} \left( p \phi^{k\beta}+q\Phi^{k\beta} \right) \approx 0\\
\bar{\mathcal{K}_i} &:= \mathcal{K}_i - \nabla_\beta \Phi_i^{\ \beta} - \varepsilon_{ijk}\,b^j_{\ \beta} \phi^{k\beta} \approx 0\\
\end{split}
\end{eqnarray}
The consistency of the secondary constraints leads to no new constraints, ending the iterative procedure here. So, we have the complete constraint structure of the theory. On examining the Poisson algebra of the constraints ({\it see \ref{App:Poisson}}), we see that this is a mixed system, with both first-class (whose algebra close with all constraints) and second-class (whose algebra does not close among themselves) constraints.

In the table below, we give a complete classification of the constraints along with an explanation of the notation. First-class constraints will be denoted as $\Sigma$ whereas second-class constraints will be denoted as $\Omega$.
\begin{table}[h]
\label{table:constraints}
\caption{Classification of Constraints}
\centering
\begin{tabular}{l  c  c}
\\[0.5ex]
\hline
\hline\\[-2ex]
 & First class $\Sigma$ & Second class $\Omega$\\[0.5ex]
\hline\\[-2ex]
Primary &\ \ $\Sigma_{(3)i}=\phi_i^{\ 0},~ \Sigma_{(4)i}=\Phi_i^{\ 0}$ &\ \ ${\Omega_{(1)}}_i^{\ \alpha}=\phi_i^{\ \alpha}, ~ {\Omega_{(2)}}_i^{\ \alpha}=\Phi_i^{\ \alpha}$\\[0.5ex]
\hline\\[-2ex]
Secondary &\ $\Sigma_{(1)i}=\bar{\mathcal{H}_i},~\Sigma_{(2)i}=\bar{\mathcal{K}_i}$ &\ \ \\[0.5ex]
\hline
\hline
\end{tabular}
\end{table}

As explained at the beginning of this section, we will now implement the method of Dirac brackets\footnote{Dirac brackets are denoted by a star $\lbrace\, , \, \rbrace^*$ to distinguish them from Poisson brackets $\lbrace\, , \,\rbrace$.}  and thus eliminate all second-class constraints from the theory. The Dirac bracket is defined in terms of Poisson brackets as,
\begin{eqnarray}
\label{Dirac defn}
\lbrace f(x),g(x') \rbrace^* := \lbrace f(x),g(x') \rbrace - \displaystyle{\sum_{(YZ)}} \int dy dz ~ \lbrace f(x),\Omega_{(Y)}(y)\rbrace ~\Delta^{-1}_{(YZ)}(y,z) ~\lbrace \Omega_{(Z)}(z), g(x') \rbrace
\end{eqnarray}
The quantity $\Delta^{-1}_{(YZ)}(y,z)$ is the inverse of the matrix $\Delta_{(YZ)}(y,z)$, formed from the second class constraints $\Omega_{(Z)}$, with $Y,Z=1,2$. The elements of the matrix $\Delta_{(YZ)}(y,z)=\lbrace \Omega_{(Y)}, \Omega_{(Z)} \rbrace$ are given by
\begin{eqnarray}
\label{const matrix}
\left[\Delta_{(YZ)}(x,x')\right]^{\ \ \alpha\beta}_{ij} = - 2 ~\varepsilon^{0\alpha\beta}~\eta_{ij} \left(\begin{array}{rr}\alpha_4 & a\\a & \alpha_3 \end{array}\right)  \delta(x-x'),
\end{eqnarray}
and the matrix $\Delta^{-1}_{(YZ)}(y,z)$ can thus be written down as
\begin{eqnarray}
\label{inv const matrix}
\left[\Delta^{-1}_{(YZ)}(x,x')\right]_{\ \ \beta\alpha}^{ij} = -\frac{1}{2\left(\alpha_3\alpha_4-a^2\right)} \varepsilon_{0\beta\alpha} ~\eta^{ij} \left(\begin{array}{rr}\alpha_3 & -a\\-a & \alpha_4 \end{array}\right)  \delta(x-x').
\end{eqnarray}
Here the condition $\alpha_3\alpha_4-a^2 \neq 0$ ensures the invertibility of the matrix $\Delta_{YZ}$. This is the same condition as encountered before, (see footnote in page \pageref{fn:neq 0 condition}) and we observe that it also comes up naturally in this canonical analysis. The Dirac brackets between pairs of basic fields and momenta can now be computed, and they \emph{all} turn out to be non-zero. These brackets are listed below,
\begin{eqnarray}
\label{basicDirac}
\begin{split}
\lbrace b^i_{\ \mu}(x),b^j_{\ \nu}(x')\rbrace^* &= \frac{\alpha_3}{2\left(\alpha_3 \alpha_4 - a^2\right)} ~\varepsilon_{0\alpha\beta} ~\delta^\alpha_\mu \delta^\beta_\nu ~\eta^{ij} ~\delta(x-x')\\
\lbrace b^i_{\ \mu}(x),\pi_j^{\ \nu}(x')\rbrace^* &= \left[\delta^\nu_\mu - ~\delta^\alpha_\mu \delta^\nu_\alpha ~\frac{\left(\alpha_3 \alpha_4 - 2a^2\right)}{2\left(\alpha_3 \alpha_4 - a^2\right)} \right] ~\delta^i_j ~\delta(x-x')\\
\lbrace b^i_{\ \mu}(x),\omega_j^{\ \nu}(x')\rbrace^* &= -\frac{a}{2\left(\alpha_3 \alpha_4 - a^2\right)} ~\varepsilon_{0\alpha\beta}  \delta^\alpha_\mu \delta^\beta_\nu ~\eta^{ij} ~\delta(x-x')\\
\lbrace b^i_{\ \mu}(x),\Pi^{\ \nu}_j(x')\rbrace^* &= \frac{\alpha_3 a}{2\left(\alpha_3 \alpha_4 - a^2\right)} ~\delta^\alpha_\mu \delta^\nu_\alpha ~\delta^i_j ~\delta(x-x')\\
\lbrace \omega^i_{\ \mu}(x),\pi^{\ \nu}_j(x')\rbrace^* &= -\frac{\alpha_4 a}{2\left(\alpha_3 \alpha_4 - a^2\right)} ~\delta^\alpha_\mu \delta^\nu_\alpha ~\delta^i_j ~\delta(x-x')\\
\lbrace \omega^i_{\ \mu}(x),\Pi^{\ \nu}_j(x')\rbrace^* &= \left[\delta^\nu_\mu - ~\delta^\alpha_\mu \delta^\nu_\alpha ~\frac{\left(\alpha_3 \alpha_4 - 2a^2\right)}{2\left(\alpha_3 \alpha_4 - a^2\right)} \right] ~\delta^i_j ~\delta(x-x')\\
\lbrace \pi_i^{\ \mu}(x),\pi^{\ \nu}_j(x')\rbrace^* &= \left[\frac{\alpha_4^2\alpha_3 + 4a^3 - 2\alpha_4a^2 - 2\alpha_4\alpha_3a}{2\left(\alpha_3 \alpha_4 - a^2\right)} \right] ~\varepsilon^{0\alpha\beta} \delta^\mu_\alpha \delta^\nu_\beta ~\eta_{ij} ~\delta(x-x')\\
\lbrace \Pi_i^{\ \mu}(x),\Pi^{\ \nu}_j(x')\rbrace^* &= \frac{\alpha_4\alpha_3^2}{2\left(\alpha_3 \alpha_4 - a^2\right)} ~\varepsilon^{0\alpha\beta} \delta^\mu_\alpha \delta^\nu_\beta ~\eta_{ij}~\delta(x-x')\\
\lbrace \pi_i^{\ \mu}(x),\Pi^{\ \nu}_j(x')\rbrace^* &= \frac{\alpha_3 \alpha_4a}{2\left(\alpha_3 \alpha_4 - a^2\right)} ~\varepsilon^{0\alpha\beta} \delta^\mu_\alpha \delta^\nu_\beta ~\eta_{ij}~\delta(x-x')\\
\end{split}
\end{eqnarray}
The Dirac brackets of the second class constraints among themselves and with all other quantities turn out to be zero, as expected. Hence these constraints can be strongly set equal to zero.

We now give the complete Dirac algebra of the system by listing only the non-trivial ones. First, the involutive algebra of the first class constraints is given,
\begin{eqnarray}
\label{constrnsDirac}
\begin{split}
\lbrace\bar{\mathcal{H}_i}(x),\bar{\mathcal{H}_j}(x')\rbrace^* &=  \varepsilon_{ijk}\left(p\,\bar{\mathcal{H}^k}+q\,\bar{\mathcal{K}^k}\right)\delta(x-x')\\
\lbrace\bar{\mathcal{K}_i}(x),\bar{\mathcal{K}_j}(x')\rbrace^* &= - \varepsilon_{ijk}\,\bar{\mathcal{K}^k}\,\delta(x-x')\\
\lbrace\bar{\mathcal{H}_i}(x),\bar{\mathcal{K}_j}(x')\rbrace^* &= - \varepsilon_{ijk}\,\bar{\mathcal{H}^k}\,\delta(x-x').\\
\end{split}
\end{eqnarray}
Next, the involutive algebra of the first class constraints with the canonical hamiltonian $H_C=\int d^2x'  ~\mathcal{H}_C(x')$ is written,
\begin{eqnarray}
\label{constrnsDiracHamilt}
\begin{split}
\lbrace H_C,\bar{\mathcal{H}_i}(x)\rbrace^* &= \left[\varepsilon_{ijk}\left\lbrace\omega^j_{\ 0}(x) - p b^j_{\ 0}(x) \right\rbrace \bar{\mathcal{H}^k}(x) - q \,\varepsilon_{ijk} b^j_{\ 0}(x) ~\bar{\mathcal{K}^k}(x)\right]\\
\lbrace H_C,\bar{\mathcal{K}_i}(x)\rbrace^* &= \left[\varepsilon_{ijk} \,b^j_{\ 0}(x) ~\bar{\mathcal{H}^k}(x) + \varepsilon_{ijk} \,\omega^j_{\ 0}(x) ~\bar{\mathcal{K}^k}(x)\right]\\
\lbrace H_C,\pi_i^{\ 0}(x)\rbrace^* &= \bar{\mathcal{H}_i}(x)\\
\lbrace H_C,\Pi_i^{\ 0}(x)\rbrace^* &= \bar{\mathcal{K}_i}(x).\\
\end{split}
\end{eqnarray}
Note that now we have a system with only first class constraints whose Dirac algebra has been given above. The second class constraints, as already stated, are strongly set equal to zero. In the next section, we will use these results to systematically find the gauge generator following \cite{Banerjee:1999hu, Banerjee:1999yc} and show that it generates \emph{off-shell} symmetries of the action (\ref{action}).

\section{Gauge symmetries of the action}

In this section, we systematically calculate the gauge symmetry generator $G$ of the action (\ref{action}). We follow the method enunciated in \cite{Banerjee:1999hu, Banerjee:1999yc} to construct $G$. It is to be noted at the very onset that this method does not require any use of the equations of motion. Consequently the generated symmetries are off-shell. This may be compared to the approach \cite{Castellani:1981us} adopted in \cite{Blagojevic:2004hj}, for discussions in this model, where the generator maps solutions to solutions of the equations of motion. Since equations of motion are involved, it becomes debatable whether the generator would be able to reproduce the genuine (off-shell) symmetries of the model. In this sense our approach is conceptually cleaner.  We next outline this approach briefly.

Having eliminated all the second class constraints through introduction of Dirac brackets, we are left with a theory with only first class constraints. The set of constraints $\Sigma_{(I)}$ is now classified as
\begin{equation}
\label{RB const1}
\left[\Sigma_{(I)}\right] = \left[\Sigma_{(A)};\Sigma_{(Z)}\right]
\end{equation}
where $A=3,4$ belong to the set of primary (first class) constraints, $Z=1,2$ to the set of secondary (first class) constraints and $I=1,2,3,4$ refer to all (first class) constraints. The total hamiltonian is
\begin{equation}
\label{RB tot hamilt}
H_{T} = H_{C} + \int d^2x ~\lambda^{(A)}\Sigma_{(A)}
\end{equation}
where $H_C$ is the canonical hamiltonian and $\lambda^{(A)}$ are Lagrange multipliers enforcing the primary constraints. The most general expression for the generator of gauge transformations is obtained according to the Dirac conjecture as
\begin{equation}
\label{RB gen G}
G = \int d^2x ~\epsilon^{(I)}\Sigma_{(I)}
\end{equation}
where $\epsilon^{(I)}$ are the gauge parameters. However, not all of these are independent.  This is most simply and elegantly seen by  demanding the commutation of an arbitrary gauge variation with the total time derivative, i.e. $\frac{d}{dt}\left(\delta q \right) = \delta \left(\frac{d}{dt} q \right)$. Recalling that,
\begin{eqnarray}
\begin{split}
\delta q = \lbrace q, G \rbrace^*\\
\frac{dq}{dt} = \lbrace q, H_T \rbrace^*,\\
\end{split}
\end{eqnarray}
a little algebra, using (\ref{RB tot hamilt} and \ref{RB gen G}), yields the following conditions \cite{Banerjee:1999hu, Banerjee:1999yc}
\begin{eqnarray}
\label{RB master 1}
\delta\lambda^{(A)}(x) \!\!&=&\!\! \displaystyle\frac{d\epsilon^{(A)}(x)}{dt}-\int d^2x' \,\epsilon^{(I)}(x') \,\left[ \left(V^A_{\;\;\: I}\right)(x,x')+\int d^2x''\,\lambda^{(B)}(x'') ~\left(C^A_{\;\;\, IB}\right)(x,x',x'')\right]\\
\label{RB master 2}
0 \!\!&=&\!\! \displaystyle\frac{d\epsilon^{(Z)}(x)}{dt} - \int d^2x'~\epsilon^{(I)}(x') \,\left[ \left(V^Z_{\;\;\: I}\right)(x,x') +\int d^2x''\,\lambda^{(B)}(x'') \,\left(C^Z_{\;\;\, IB}\right)(x,x',x'')\right].
\end{eqnarray}
Here the coefficients $\left(V^I_{\;\;\: J}\right)(x,x')$ and $\left(C^I_{\;\;\, JK}\right)(x,x',x'')$ are the structure functions of the involutive (first-class) algebra, defined through
\begin{eqnarray}
\label{StructureConsts}
\begin{split}
\lbrace\Sigma_{(I) i}(x),\Sigma_{(J) j}(x')\rbrace^* = &\int d^2x''\,{\left(C^K_{\;\;\, IJ}\right)}_{ijk}(x'',x,x') ~\Sigma_{(K)}^{\quad k}(x'')\\
\lbrace H_C,\Sigma_{(I)i}(x)\rbrace^* = &\int d^2x'\,{\left(V^J_{\;\;\: I}\right)}_{ik}(x',x)~\Sigma_{(J)}^{\quad k}(x').\\
\end{split}
\end{eqnarray}
The second condition\footnote{The significance of the other condition (\ref{RB master 1}) is discussed in \ref{App:2nd Master}.}  (\ref{RB master 2}) makes it is possible to choose $A$ independent gauge parameters from the set $\epsilon^{(I)}$ and express the generator $G$ of (\ref{RB gen G}) entirely in terms of them. This shows that the number of independent gauge parameters is equal to the number of independent, primary first-class constraints. 

Before proceeding further let us note the following point. The derivation of (\ref{RB master 2}) is based only on the relation between the velocities and the canonical momenta, namely, the first of Hamilton's equations \cite{Banerjee:1999hu, Banerjee:1999yc}. Note that the full dynamics, implemented through the second of Hamilton's equations $\left(\frac{dp}{dt}=\lbrace p, H \rbrace\right)$ is not required to impose restrictions on the gauge parameters. Since this is the only input in our method of abstraction of the independent gauge parameters, we find that our analysis will be valid off-shell. The off-shell invariance will also be demonstrated explicitly.

The structure constants defined in (\ref{StructureConsts}) can now be obtained using the results of the various Dirac brackets (\ref{basicDirac}, \ref{constrnsDirac} \& \ref{constrnsDiracHamilt}). These are:
\begin{eqnarray}
\label{Cs}
\begin{split}
{\left(C^1_{\;\; 11}\right)}_{ijk}(x'',x,x') &= p \,\varepsilon_{ijk} \,\delta(x-x'') \delta(x''-x')\\
{\left(C^2_{\;\; 11}\right)}_{ijk}(x'',x,x') &= q \,\varepsilon_{ijk} \,\delta(x-x'') \delta(x''-x')\\
{\left(C^1_{\;\; 12}\right)}_{ijk}(x'',x,x') &= - \varepsilon_{ijk} \,\delta(x-x'') \delta(x''-x')\\
{\left(C^2_{\;\; 12}\right)}_{ijk}(x'',x,x') &= 0\\
{\left(C^1_{\;\; 22}\right)}_{ijk}(x'',x,x') &= 0\\
{\left(C^2_{\;\; 22}\right)}_{ijk}(x'',x,x') &= - \varepsilon_{ijk} \,\delta(x-x'') \delta(x''-x')\\
{\left(C^A_{\;\; IJ}\right)}_{ijk}(x'',x,x') &= 0.\\
{\left(C^Z_{\;\; AB}\right)}_{ijk}(x'',x,x') &= 0.\\
\end{split}
\end{eqnarray}
and,
\begin{eqnarray}
\label{Vs}
\begin{split}
{\left(V^1_{\;\;\, 1}\right)}_{ik}(x',x) &= \varepsilon_{ijk} \,\left[\omega^j_{\ 0}(x') - p\,b^j_{\ 0}(x')\right] \delta(x-x')\\
{\left(V^2_{\;\;\, 1}\right)}_{ik}(x',x) &= -q \,\varepsilon_{ijk} \,b^j_{\ 0}(x') \,\delta(x-x')\\
{\left(V^1_{\;\;\, 2}\right)}_{ik}(x',x) &= \varepsilon_{ijk} \,b^j_{\ 0}(x') \,\delta(x-x')\\
{\left(V^2_{\;\;\, 2}\right)}_{ik}(x',x) &= \varepsilon_{ijk} \,\omega^j_{\ 0}(x') \,\delta(x-x')\\
{\left(V^1_{\;\;\, 3}\right)}_{ik}(x',x) &= \eta_{ik} \,\delta(x-x')\\
{\left(V^2_{\;\;\, 3}\right)}_{ik}(x',x) &= 0\\
{\left(V^1_{\;\;\, 4}\right)}_{ik}(x',x) &= 0\\
{\left(V^2_{\;\;\, 4}\right)}_{ik}(x',x) &= \eta_{ik} \,\delta(x-x').\\
{\left(V^A_{\;\;\, I}\right)}_{ik}(x',x) &= 0
\end{split}
\end{eqnarray}
Now the generator (\ref{RB gen G}) is expanded as,
\begin{eqnarray}
\label{generatorGen}
G=\int d^2x \left[\epsilon^{(3)i}(x)\,\pi^{\ 0}_i(x)+\epsilon^{(4)i}(x)\,\Pi^{\ 0}_i(x) + \epsilon^{(1)i}(x)\,\bar{\mathcal{H}_i}(x) + \epsilon^{(2)i}(x)\,\bar{\mathcal{K}_i}(x)\right]
\end{eqnarray}
where the parameters $\epsilon^{(I)i}$ are not all independent, but satisfy the equation (\ref{RB master 2}), so that,
\begin{eqnarray}
\label{master eq}
\displaystyle\frac{d\epsilon^{\scriptscriptstyle{(Z)i}}(x)}{dt}-\int d^2x' \,\epsilon^{\scriptscriptstyle{(I)k}}(x')\,{\left({V^{\scriptscriptstyle{Z}}}_{\scriptscriptstyle{\,I}}\right)}_k^{\;\;\,i}(x',x)=0
\end{eqnarray}
Using the structure constants ${\left({V^{\scriptscriptstyle{Z}}}_{\scriptscriptstyle{\,I}}\right)}_k^{\;\;\,i}(x',x)$ already determined in (\ref{Vs}), we get the following two relations among the parameters $\epsilon^{\scriptscriptstyle{(Z)}}$
\begin{eqnarray}
\label{rel epsilons}
\begin{split}
\dot{\epsilon}^{(1)i}(x) &= \epsilon^{(3)i}(x) + \epsilon^{(1)k}(x) \,\varepsilon_k^{\;\:ij}\left[ p\,b_{j0}(x)-\omega_{j0}(x) \right] - \epsilon^{(2)k}(x) \,\varepsilon_k^{\;\:ij}\,b_{j0}(x)\\
\dot{\epsilon}^{(2)i}(x) &= \epsilon^{(4)i}(x) + q \,\epsilon^{(1)k}(x) \,\varepsilon_k^{\;\: ij}\,b_{j0}(x) - \epsilon^{(2)k}(x) \,\varepsilon_k^{\;\: ij}\,\omega_{j0}(x).\\
\end{split}
\end{eqnarray}
After using these equations (\ref{rel epsilons}) in the generator (\ref{generatorGen}) to eliminate the gauge parameters $\epsilon^{(3)}$ and $\epsilon^{(4)}$, we obtain our cherished structure in terms of the two independent gauge parameters $\epsilon^{(1)}$ and $\epsilon^{(2)}$,
\begin{eqnarray}
\label{generatorOur}
G = \int d^2x \!\!\!\!\!\!&&\!\!\!\!\!\! \left[\left\lbrace \dot{\epsilon}^{(1)i}(x) -  \epsilon^{(1)k}(x) \,\varepsilon_k^{\;\: ij}\,\left[ p\,b_{j0}(x)-\omega_{j0}(x) \right] + \epsilon^{(2)k}(x) \,\varepsilon_k^{\;\: ij}\,b_{j0}(x)\right\rbrace\,\pi_i^{\ 0}(x) \right.\nonumber\\
&+&\!\!\!\!\left.\left\lbrace \dot{\epsilon}^{(2)i}(x) - q \,\epsilon^{(1)k}(x) \,\varepsilon_k^{\;\: ij}\,b_{j0}(x) + \epsilon^{(2)k}(x) \varepsilon_k^{\;\: ij}\,\omega_{j0}(x) \right\rbrace \Pi_i^{\ 0}(x)  \right.\nonumber\\
&+&\!\!\!\left. \epsilon^{(1)i}(x)\,\bar{\mathcal{H}_i}(x) + \epsilon^{(2)i}(x)\,\bar{\mathcal{K}_i}(x)\right].
\end{eqnarray}
On rearranging the generator and renaming the parameters as $\epsilon^{(1)}=\epsilon$ and $\epsilon^{(2)}=\tau$, we obtain the generator in the form
\begin{eqnarray}
\label{generatorFinal}
\begin{split}
G=&\int d^2x \left[\mathcal{G}_\epsilon(x)+\mathcal{G}_\tau(x)\right]\\
&\mathcal{G}_\epsilon=\dot{\epsilon}^i\,\pi_i^{\ 0} + \epsilon^i\left[\bar{\mathcal{H}_i}- \varepsilon_{ijk} \big( \omega^j_{\ 0} - p\,b^j_{\ 0}\big)\pi^{k0} + q \,\varepsilon_{ijk}\,b^j_{\ 0}\Pi^{k0} \right]\\
&\mathcal{G}_\tau=\dot{\tau}^i\Pi_i^{\ 0} + \tau^i\left[\bar{\mathcal{K}_i}-\varepsilon_{ijk}\big(b^j_{\ 0}\,\pi^{k0} + \omega^j_{\ 0}\,\Pi^{k0}\big)\right]\\
\end{split}
\end{eqnarray}

The generator thus written gives rise to gauge variations of fields in the theory. The transformations of the basic fields $b^i_{\ \mu}$ and $\omega^i_{\ \mu}$ are:
\begin{eqnarray}
\label{field transf gauge}
\begin{split}
\delta b^i_{\ \mu}(x) &:= \lbrace b^i_{\ \mu}(x),G \rbrace^* = \partial_\mu\epsilon^i(x) + \varepsilon^i_{\ jk} \,\omega^j_{\ \mu}(x) \epsilon^k(x) - p \,\varepsilon^i_{\ jk} \,b^j_{\ \mu}(x) \epsilon^k(x) + \varepsilon^i_{\ jk}\,b^j_{\ \mu}(x) \tau^k(x),\\
\delta \omega^i_{\ \mu}(x) &:= \lbrace \omega^i_{\ \mu}(x),G \rbrace^* = \partial_\mu \tau^i(x) + \varepsilon^i_{\ jk} \,\omega^j_{\ \mu}(x) \tau^k(x) - q \,\varepsilon^i_{\ jk} \,b^j_{\ \mu}(x) \epsilon^k(x).\\
\end{split}
\end{eqnarray}    

We would now like to demonstrate the explicit \emph{off-shell} invariance of the action (\ref{action}) under the above gauge transformations of the fields (\ref{field transf gauge}). The variation of the action, in general, reads:
\begin{eqnarray}
\label{gen var action}
\delta I \ =\  \delta I\Big|_\text{Einstein} + \delta I\Big|_\text{Cosmological} + \delta I\Big|_\text{Chern Simons} + \delta I\Big|_\text{Torsion}
\end{eqnarray}
Substituting our gauge transformations in the above, we observe that $\delta I$ vanishes without using any equations of motion. The cancellation of relevant terms is quite interesting and we would like to note certain features involved. An easy way of seeing the cancellation is to begin by focusing on families of similar structured terms. For example, terms containing one derivative, the parameter $\epsilon$, $b$ and $\omega$ (where the indices have been suppressed for simplicity) may occur as $\left(~\varepsilon^{\mu\nu\rho} \varepsilon^i_{\ jk} ~b^j_{\ \mu} ~\partial_\nu \omega_{i\rho} ~\epsilon^k ~\right)$ or $\left(~\varepsilon^{\mu\nu\rho} \varepsilon^i_{\ jk} ~\partial_\mu \epsilon_i ~\omega^j_{\ \nu} ~b^k_{\ \rho} ~\right)$ or in other such different types. However all of them may be cast as the same term on using the properties of the levi-civita symbols and/or using partial integrals.  When all such families are identified, we see that there occur two different types of cancellation. First, many terms are identically zero or cancel algebraically, needing at most throwing of some total derivatives. Secondly, in some cases terms from different pieces of the action, with their different parameters, cancel by virtue of the relation between parameters $a, \Lambda, \alpha_3\, \&\, \alpha_4$ and the definition of the quantities `$p$' and `$q$'. We now demonstrate this for one particular family. The terms containing $(\omega\,b\ b~\epsilon)$ can be collected from the variations of different pieces of the action (\ref{action}). These are written below in exactly the same order as they appear in (\ref{gen var action}),
\begin{eqnarray}
\delta I\Big|_{(\omega\,b\ b\,\epsilon) \,\text{terms}} &=& -2aq \int d^2x \,\varepsilon^{\mu\nu\rho} \varepsilon_{ijk} \varepsilon^j_{\ lm}\, b^i_{\ \mu} \omega^k_{\ \rho} b^l_{\ \nu} \,\epsilon^m - 2\Lambda \int d^2x \,\varepsilon^{\mu\nu\rho} \,\omega^j_{\ \mu} b_{j\nu} b_{k\rho} \,\epsilon^k \nonumber\\
&\,& +\, ~0~ + ~\alpha_4 p \int d^2x ~\varepsilon^{\mu\nu\rho} \left[\varepsilon^m_{\ \ il} \varepsilon_{mpq} \,b^i_{\ \mu} \omega^l_{\ \nu} b^p_{\ \rho} \,\epsilon^q + \varepsilon^i_{\ jk} \varepsilon_{ilm} \,b^j_{\ \mu} b^m_{\ \rho} \omega^l_{\ \nu} \,\epsilon^k \right] \nonumber\\
&=&  2\left(-\Lambda + aq +\alpha_4 p\right) \int d^2x \,\varepsilon^{\mu\nu\rho} \,\omega^j_{\ \mu} b_{j\nu} b_{k\rho} \,\epsilon^k\nonumber\\
&=& 0
\end{eqnarray}
where use of the definitions $p=\dfrac{\alpha_3\Lambda + \alpha_4 a}{\alpha_3\alpha_4 - a^2}$ and $q=-\dfrac{\alpha_4^{\ 2} +a\Lambda}{\alpha_3\alpha_4 - a^2}$ has been made, to observe that the combination $\left(-\Lambda + aq +\alpha_4 p\right)=0$.
A summary of the different terms and their cancellation factors are given below. Terms that are not explicitly mentioned here, reduce to zero algebraically.
\begin{table}[h]
\label{table:zero combs}
\caption{Cancellation of families of terms}
\centering
\begin{tabular}{l  c}
\\[-0.5ex]
\hline
Term & Combination of parameters giving zero\\[0.5ex]
\hline
$\omega\,b\ b~\epsilon$ & $-\Lambda + aq +\alpha_4 p$ \\[0.5ex]
$\omega\,\omega\ b~\epsilon$ & $\alpha_4 + q \alpha_3 + ap$ \\[0.5ex]
\hline
\end{tabular}
\end{table}

We are thus led to an intriguing situation. There are two sets of field transformations, one derived above in (\ref{field transf gauge}), and the other from the Poincare gauge gravity (\ref{fieldtrans3D}), both of these being true symmetries of the action. We have explicitly demonstrated this by showing that variations in the action (\ref{action}), under any of the two transformations, vanish without requiring any use of the equations of motion. Consequently these are proper gauge symmetries, i.e. they are off-shell symmetries. One would therefore expect an off-shell mapping between the two sets of parameters $\left(\epsilon,~\tau\right)$ and $\left(\xi,~\theta\right)$. Alas, this map does not exist. Indeed, the following map, which was also mentioned in the literature \cite{Blagojevic:2004hj},
\begin{eqnarray}
\label{onshell map}
\begin{split}
\epsilon^i &= -\xi^\lambda \,b^i_{\ \lambda}\\
\tau^i &= -\left(\theta^i + \xi^\lambda \,\omega^i_{\ \lambda} \right).\\
\end{split}
\end{eqnarray}
connects the two transformations by the identification,
\begin{eqnarray}
\label{onshell relation transfs}
\begin{split}
\delta_0 b^i_{\ \mu} &= \delta_{PGT} b^i_{\ \mu} - \xi^\rho\left(T^i_{\ \mu\rho}-p \,\varepsilon^{ijk}\, b_{j\mu} b_{k\rho} \right)\\
\delta_0 \omega^i_{\ \mu} &= \delta_{PGT} \omega^i_{\ \mu} - \xi^\rho\left(R^i_{\ \mu\rho}-q \,\varepsilon^{ijk}\, b_{j\mu} b_{k\rho} \right).\\
\end{split}
\end{eqnarray}
The terms within parentheses, which are exactly the terms destroying the mapping between $\delta_0$ and $\delta_{PGT}$, are the equations of motion (\ref{eqnmot}). So the map (\ref{onshell map}) holds only \emph{on-shell}. In the next section we attempt towards a possible understanding of this point.

\section{Comments on the lack of off-shell mapping between the transformation parameters of PGT and the independent gauge parameters}

  It has been observed in the last section that the gauge transformations (\ref{field transf gauge}) of the basic fields of the theory (\ref{action}) cannot be mapped on the transformations of the same under PGT, namely, (\ref{fieldtrans3D}) {\it{without invoking}} the equations of motion, although both sets of transformations preserve the off-shell invariance of the same action. Stated otherwise, we don't have an off-shell mapping between the two sets of parameters characterising these transformations. This, notwithstanding the facts that the number of independent parameters of the two sets match exactly and both the sets provide off-shell invariance of (\ref{action}) as we have explicitly demonstrated above. We show  that this feature is a peculiarity of the PGT framework.

Before considering PGT, let us first analyse metric gravity theory in the second order formalism, given by the action 
\begin{eqnarray}
I=\int ~d^3x~ \sqrt{-g} R
\label{Einsteinaction}
\end{eqnarray}
where $R$ is the Ricci scalar. Here $g = {\rm{det}}g_{\mu\nu}$ , $g_{\mu\nu}$ being the metric tensor. The theory is invariant under diffeomorphism
\begin{equation}
x^\mu \to x^\mu + \xi^\mu\label{Diff}
\end{equation}
The canonical analysis of the gauge transformations of the theory following the method of \cite{Banerjee:1999hu,Banerjee:1999yc} was performed in \cite{Mukherjee:2007yi}. The analysis was done in (3+1) dimensions but it can be easily adapted to (2+1) dimensions which is relevant here. For the canonical analysis, spacetime is foliated in spacelike two-surfaces as per the Arnowit-Deser-Misner (ADM) decomposition. The lapse variable $N^{\bot}$ represents an arbitrary variation normal to the two-surface on which the state of the system is defined whereas the shift variables $N^{\alpha}$  represent variations along the surface. They are defined by{\footnote{Note that $g^{\alpha\beta}$ is the inverse of the spatial metric $g_{\alpha\beta}$ on the two surface.}}
\begin{eqnarray}
N^{\beta} &=& g^{\alpha\beta}g_{0\alpha}
\label{Nj}
\end{eqnarray}
\begin{eqnarray}
N^{\perp} &=& \left(- g^{00}\right)^{-1/2} 
\label{N} 
\end{eqnarray}
These variables are not really the dynamical variables of the theory. By adding suitable divergences to the action (\ref{Einsteinaction}) we can write an equivalent lagrangian \cite{Hanson:1976cn, Sundermeyer:1982}
\begin{eqnarray}
\int d^{2}x {\cal{L}} =  \int d^{2}x N^{\perp} \left(g\right) ^{1/2} \left(K_{\alpha\beta}K^{\alpha\beta} - K^{2} + R \right) 
\label{L} 
\end{eqnarray}
where $ K = K^{\alpha}{}_{\alpha} = g^{\alpha\beta} K_{\alpha\beta}$ and $R$ is the Ricci scalar on the two surface. The second fundamental form $K_{\alpha\beta}$ is defined as 
\begin{eqnarray}
K_{\alpha\beta} &=& \frac{1}{2 N^{\bot}} \left(- \dot{g}_{\alpha\beta} + N_{\alpha \mid \beta } + N_{\beta \mid \alpha} \right)
\label{K} 
\end{eqnarray}
The ${\mid}$ indicates covariant derivative on the two-surface. The lagrangian (\ref{L}) is suitable for canonical analysis because it does not contain the time derivatives of the lapse and shift variables i.e. in the canonical analysis they appear as Lagrange multipliers. Their conjugate momenta $\pi_{0}$ and $\pi_{\alpha}$ vanish weakly, providing the following primary constraints of the theory:
\begin{eqnarray}
\Omega_{0} = \pi_{0}  \approx 0\\
\Omega_{\alpha}  = \pi_{\alpha} \approx 0
\label{SCnew} 
\end{eqnarray}

The basic fields are $g_{\alpha\beta}$ with their conjugate momenta $\pi^{\alpha\beta}$. The canonical hamiltonian can be worked out as
\begin{eqnarray}
H_{c} &=&  \int d^{2}x \left(\pi_{\mu} \dot{N}^{\mu} + \pi^{\alpha\beta} \dot{g}_{\alpha\beta} - {\cal{L}}\right) \nonumber\\
&=& \int d^{2}x \left( N^{\perp}{\cal{H}}_{\perp} + N^{\alpha}{\cal{H}}_{\alpha} \right)
\label{canonHamiltMS}
\end{eqnarray}
where,
\begin{eqnarray}
{\cal{H}}_{\perp} &=&  g ^{-1/2} \left( \pi_{\alpha\beta}\pi^{\alpha\beta} - \frac{1}{2} \pi^{2}\right) - \left(g\right) ^{1/2}R\\
{\cal{H}}_{\alpha} &=& - 2 \pi_{\alpha}{}^{\beta}{}_{\mid \beta }.
\label{H2} 
\end{eqnarray}
The total hamiltonian is given by,
\begin{eqnarray}
H_T = H_c + \int d^2x ~\left[\lambda^0 \Omega_0 + \lambda^\alpha\Omega_\alpha \right]
\label{HTMS}
\end{eqnarray}
where $\lambda^0$, $\lambda^\alpha$ are multipliers enforcing the primary constraints $\Omega_0$, $\Omega_\alpha$. The secondary constraints, found by time conserving the primary constraints, are
\begin{eqnarray}
\Omega_{3} = \lbrace \pi_0, H_T \rbrace = {\cal{H}}_{\perp} \approx 0\\
\Omega_{3+\alpha} = \lbrace \pi_\alpha, H_T \rbrace = {\cal{H}}_{\alpha} \approx 0.
\label{SC} 
\end{eqnarray}
No further constraints are generated by this iterative procedure. Note that all the constraints are first class. So, following Dirac's hypothesis \cite{Dirac:Lectures}, the gauge generator can be written as
\begin{eqnarray}
G = \int d^{2}x \left(\epsilon^{0}\Omega_{0} + \epsilon^{\alpha}\Omega_{\alpha} + \epsilon^{3}\Omega_{3} + \epsilon^{3+\alpha}\Omega_{3+\alpha}\right),
\label{GG} 
\end{eqnarray}
where $\epsilon^0$, $\epsilon^\alpha$, $\epsilon^3$ and $\epsilon^{3+\alpha}$ are gauge parameters. Now using our master equation (\ref{RB master 2}) we get \cite{Mukherjee:2007yi}
\begin{eqnarray}
\epsilon^{0} \left(x\right) &=&\left[  \dot{\epsilon}^{3} + \epsilon^{3+\alpha} \partial_{\alpha}N^{\bot} - N^{\alpha} \partial_{\alpha}\epsilon^{3}\right] \left(x\right)\\
\epsilon^{\alpha} \left(x\right)&=&\left[  \dot{\epsilon}^{3+\alpha} + \epsilon^{3+\beta} \partial_{\beta}N^{\alpha} - N^{\beta}\partial_{\beta} \epsilon^{3+\alpha} - N^{\bot}g^{\beta\alpha}\partial_{\beta}\epsilon^{3} + \epsilon^{3}g^{\beta\alpha}\partial_{\beta}N^{\bot}\right] \left(x\right),
\label{IGP} 
\end{eqnarray}
which shows that only three gauge parameters $\left( \epsilon^{3}, \epsilon^{3+\alpha}\right)$ are independent. Their number is equal to the number of primary first class constraints, in conformity with the discussion below (\ref{StructureConsts}). Also, this number matches with the number of diffeomorphism parameters $\xi^{\mu}$ (see \ref{Diff}).

The mapping between the gauge and diff parameters is now found by comparing the variations of $N^\perp$, $N^\alpha$ and $g_{\alpha\beta}$ under both these symmetry operations. First, we consider the gauge variations which are found by Poisson bracketing with the generator, 
\begin{eqnarray}
\delta N^{\bot}\left(x \right) = \lbrace N^\perp(x),G \rbrace = \left[  \dot{\epsilon}^{3} + \epsilon^{3+\alpha} \partial_{\alpha}N^{\bot} - N^{\alpha} \partial_{\alpha}\epsilon^{3}\right] \left(x\right)
\label{GV1}
\end{eqnarray}
\begin{eqnarray}
\delta N^{\alpha}\left(x \right) & = & \left\{N^{\alpha}\left(x\right), G\right\}\nonumber\\  &=& \left[  \dot{\epsilon}^{3+\alpha} + \epsilon^{3+\beta} \partial_{\beta}N^{\alpha} - N^{\beta}\partial_{\beta} \epsilon^{3+\alpha} - N^{\bot}g^{\beta\alpha}\partial_{\beta}\epsilon^{3} + \epsilon^{3}g^{\beta\alpha}\partial_{\beta}N^{\bot}\right] \left(x\right)
\label{GV2}
\end{eqnarray}
\begin{eqnarray}
\delta g_{\alpha\beta} \left( x \right)  & = &  \left\lbrace g_{\alpha\beta} \left( x \right), G \right\rbrace \nonumber\\ 
& = & -2 \epsilon^{3} K_{\alpha\beta} +  \epsilon^{3+\gamma} \partial_{\gamma}g_{\alpha\beta} + g_{\gamma\alpha}\partial_{\beta} \epsilon^{3+\gamma} + g_{\gamma\beta}\partial_{\alpha} \epsilon^{3+\gamma} 
\label{gGV}
\end{eqnarray}
The variation under general coordinate transformations or diff can be worked out after a bit of calculation \cite{Mukherjee:2007yi}. The desired variations are:
\begin{eqnarray}
\delta N^{\bot}\left(x \right) &=& \left(\frac{d}{dt} - N^{\alpha}\partial_{\alpha}\right)\xi^{0}N^{\bot}+\xi^{0}N^{\alpha}\partial_{\alpha}N^{\bot} + \xi^{\alpha}\partial_{\alpha}N^{\bot} \label{R1} \\
\delta N^{\alpha}\left(x \right) &=& \left(\frac{d}{dt} - N^{\beta}\partial_{\beta} \right) \left( \xi^{\alpha} + \xi^{0}N^{\alpha}\right)+ \left(\xi^{\beta} + \xi^{0}N^{\beta}\right) \partial_{\beta}N^{\alpha} - \left( N^{\bot} \right)^{2} g^{\alpha\beta} \partial_{\beta} \xi^{0} \label{R2}\\
\delta g_{\alpha\beta} \left( x \right)  &=&  \left( \xi^{0} \frac{d}{dt}  - \xi^{\gamma} \partial_{\gamma} \right) g_{\alpha\beta}  + N_{\alpha} \partial_{\beta} \xi^{0} + N_{\beta} \partial_{\alpha} \xi^{0} + g_{\gamma\alpha}\partial_{\beta} \xi^{\gamma} + g_{\gamma\beta}\partial_{\alpha} \xi^{\gamma} \label{gR}
\end{eqnarray}
Now comparing, for instance, (\ref{GV2}) and (\ref{R2}), we can establish the mapping between the independent gauge and diffeomorphism parameters as 
\begin{eqnarray}
\epsilon^{3+\alpha} & = & \xi^{\alpha} +  \xi^{0}N^{\alpha}
\label{Munattended}
\end{eqnarray}
\begin{eqnarray}
\epsilon^{3} & = & N^{\bot}\xi^{0}
\label{M}
\end{eqnarray}
This mapping, when substituted in the gauge variations of the basic fields $N^{\bot}$ and $g_{\alpha\beta}$ (equations (\ref{GV1}) and (\ref{gGV}) respectively) transforms them identically to their corresponding reparametrization variations i.e. equations (\ref{R1}) and (\ref{gR}). The equivalence of the gauge and diffeomorphism symmetries is thus established.
Note that this is an off-shell equivalence and the equations of motion are at no point invoked to establish it.

We now focus our attention on (\ref{action}). By chosing $a = 1$ and $\Lambda=\alpha_3=\alpha_4=0$ (for which $p=q=0$), it reduces to
\begin{eqnarray}
\label{FEinsteinaction}
 I=\int \,d^3x\, \varepsilon^{\mu\nu\rho} \,b^i_{\ \mu} \,R_{i\nu\rho}.
\end{eqnarray}
This is equivalent to (\ref{Einsteinaction}) as can be verified by using the identity (\ref{theidentity}) and the relation (\ref{metric}). An intriguing exercise will be to compare the gauge variations and the PGT variations for the theory (\ref{FEinsteinaction}). Referring back to (\ref{field transf gauge}) and noting $p=q=0$, the gauge variations read:
\begin{eqnarray}
\label{field transf gauge new}
\begin{split}
\delta b^i_{\ \mu}(x)  &= \partial_\mu\epsilon^i(x) + \varepsilon^i_{\ jk} \,\omega^j_{\ \mu}(x) \,\epsilon^k(x) + \varepsilon^i_{\ jk}\,b^j_{\ \mu}(x) \,\tau^k(x),\\
\delta \omega^i_{\ \mu}(x)  &= \partial_\mu \tau^i(x) + \varepsilon^i_{\ jk}\,\omega^j_{\ \mu}(x) \,\tau^k(x).\\
\end{split}
\end{eqnarray}
Comparison with the PGT transformations (\ref{fieldtrans3D}) shows that there is still no off-shell correspondence between the two transformations. Clearly, the PGT framework is distinct from the conventional one as far as the treatment of symmetries is concerned.

A similar manifestation of the same phenomenon occurs in the interpretation of 3-dimensional gravity (\ref{FEinsteinaction})  as a Chern-Simons gauge theory \cite{Witten:1988hc}. The isometry group of $M_3$ is the Poincare group $P(1,2)$. If we consider this as an ordinary gauge theory  a general gauge transformation is written as \cite{Blagojevic:2002du}
\begin{equation}
u = -\epsilon^i P_i - \tau^i J_i
\label{g1}
\end{equation}
where $P_i$ and $J_i$ are the generators of the gauge group and $\epsilon^i$ and $\tau^i$ are the gauge parameters. Introduce the corresponding gauge potential
\begin{equation}
A_\mu = b^i_{\ \mu} P_i + \omega^i_{\ \mu} J_i
\label{g2}
\end{equation} 
The variation of $A_\mu$ under a gauge transformation parametrised by (\ref{g1}) is given by 
\begin{equation}
\delta A_\mu = -\partial_\mu u - \left[A_\mu,u\right]
\label{g3}
\end{equation} 
The field strength is defined in the usual way
\begin{equation}
F_{\mu\nu} := \left[\nabla_\mu, \nabla_{\nu} \right]=\partial_\mu A_\nu - \partial_\nu A_\mu
              +\left[A_\mu ,A_\nu\right]\label{gv1}
\end{equation}
Using the explicit form of $A_\mu$ from (\ref{g2}) we find
\begin{equation}
F_{\mu\nu} = P_i \,T^i_{\ \mu\nu} + J_i \,R^i_{\ \mu\nu}\label{gv2}
\end{equation}
where the expressions of $T^i_{\ \mu\nu}$ and $R^i_{\ \mu\nu}$ coincide with (\ref{pgt}) if we identify $b^i_{\ \mu}$ and $\omega^i_{\ \mu}$ with the corresponding PGT fields. The correspondence of the Poincare gauge theory with an ordinary Chern-Simons gauge theory is so far exact.

  At this point one naturally enquires about the gauge transformations of $b^i_{\ \mu}$ and $\omega^i_{\ \mu}$. From (\ref{g1} - \ref{g3}) we get
\begin{eqnarray}
\label{g4}
\begin{split}
\delta b^i_{\ \mu}(x) & = \partial_\mu\epsilon^i(x) + \varepsilon^i_{\ jk} \,\omega^j_{\ \mu}(x) \epsilon^k(x) + \varepsilon^i_{\ jk}\,b^j_{\ \mu}(x) \tau^k(x),\\
\delta \omega^i_{\ \mu}(x) & = \partial_\mu \tau^i(x) + \varepsilon^i_{\ jk}\,\omega^j_{\ \mu}(x) \tau^k(x) .\\
\end{split}
\end{eqnarray} 
which are the same transformations as (\ref{field transf gauge new}). These are naturally invariances of (\ref{FEinsteinaction}). They, however do not map off-shell to the transformations (\ref{fieldtrans}). Note further that the latter transformations are also invariances of the same action (\ref{FEinsteinaction}). So the Poincare gauge transformations are independent of the gauge transformations of the Poincare group.

\subsection{A lagrangian analysis:}

In this subsection we will provide a lagrangian based analysis of symmetries. This will further elucidate the mismatch between Poincare gauge transformations and the standard gauge transformations.

We begin with the familiar example of electromagnetism. The action is,
\begin{equation}
S = \int {{L}}(A_\mu,\partial_\nu A_\mu) = \int F^2
\end{equation}
where $F$ is the electromagnetic field tensor. The action is invariant under the gauge transformation $A_\mu \to A_\mu + \partial_\mu \Lambda$, where $\Lambda$ is the gauge transformation parameter. By Taylor expansion
\begin{equation}
S[A_\mu + \partial_\mu \Lambda] = S[A_\mu] +\int \partial_\mu\Lambda \frac{\delta S}{\delta A_\mu}= S[A_\mu] -\int \Lambda \partial_\mu\frac{\delta S}{\delta A_\mu}
\end{equation}
The invariance condition $S[A_\mu + \partial_\mu \Lambda] = S[A_\mu]$ leads to the gauge identity $$\partial_\mu\frac{\delta S}{\delta A_\mu} = \partial_\mu \partial_\nu ~F^{\nu\mu} = 0.$$ Note that this holds {\it{off-shell}}. In fact if we invoke the equation of motion the gauge identity becomes a trivial $0 = 0$ statement. Note further that such a gauge identity exists corresponding to each independent gauge parameter.

Our course of action is now clear. We will write the identities corresponding to the Poincare gauge transformations (\ref{fieldtrans3D}) following from the invariance of (\ref{action}). By Taylor expansion we get 
\begin{eqnarray}
\label{PGT taylor}
&{\!}&\!\!\!\! S\left[b^i_{\ \mu} , \omega^i_{\ \mu}\right] = S\left[b^i_{\ \mu} + \delta_{\scriptscriptstyle PGT}b^i_{\ \mu}, \omega^i_{\ \mu} + \delta_{\scriptscriptstyle PGT} \omega^i_{\ \mu}\right]\nonumber\\
\Rightarrow &{\!}&\!\!\!\! S\left[b^i_{\ \mu} , \omega^i_{\ \mu}\right] = S\left[b^i_{\ \mu} -\left(\varepsilon^i_{\ jk} b^j_{\ \mu} \theta^k + \partial_\mu \xi^\lambda b^i_{\ \lambda} + \xi^\lambda \partial_\lambda b^i_{\ \mu} \right),\right.\nonumber\\
&{\!}&\left. \qquad\qquad\qquad\;\;\, \omega^i_{\ \mu} -\left( \partial_\mu \theta^i + \varepsilon^i_{\ jk} \omega^j_{\ \mu} \theta^k + \partial_\mu \xi^\lambda \omega^i_{\ \lambda} + \xi^\lambda \partial_\lambda \omega^i_{\ \mu} \right)\right]\nonumber\\
&{\!}& \qquad \qquad \  = S\left[b^i_{\ \mu} , \omega^i_{\ \mu}\right] - \int d^3x ~\frac{\delta S}{\delta b^i_{\ \mu}} ~\left(\varepsilon^i_{\ jk}b^j_{\ \mu}\theta^k + \partial_\mu \xi^\lambda b^i_{\ \lambda} + \xi^\lambda \partial_\lambda b^i_{\ \mu} \right)\nonumber\\
&{\!}& \qquad \qquad \qquad \ - \int d^3x ~\frac{\delta S}{\delta \omega^i_{\ \mu}} ~\left( \partial_\mu \theta^i + \varepsilon^i_{\ jk} \omega^j_{\ \mu} \theta^k + \partial_\mu \xi^\lambda \omega^i_{\ \lambda} + \xi^\lambda \partial_\lambda \omega^i_{\ \mu} \right)\nonumber\\
\Rightarrow &{\ }& \!\!\!\! \int d^3x \left[\frac{\delta S}{\delta b^i_{\ \mu}} \varepsilon^i_{\ jk} b^j_{\ \mu} + \frac{\delta S}{\delta \omega^i_{\ \mu}} \varepsilon^i_{\ jk} \omega^j_{\ \mu} -\partial_\mu\left(\frac{\delta S}{\delta \omega^k_{\ \mu}}\right) \right] \theta^k \nonumber\\
&{\!}& \qquad + \int d^3x \left[\frac{\delta S}{\delta b^i_{\ \mu}} \partial_\lambda b^i_{\ \mu} + \frac{\delta S}{\delta \omega^i_{\ \mu}} \partial_\lambda \omega^i_{\ \mu} - \partial_\mu \left(b^i_{\ \lambda}\frac{\delta S}{\delta b^i_{\ \mu}} + \omega^i_{\ \lambda} \frac{\delta S}{\delta \omega^i_{\ \mu}} \right) \right] \xi^\lambda = 0.
\end{eqnarray}
So the two independent identities turn out to be:
\begin{eqnarray}
\label{gauge ident PGT}
\begin{split}
&{\ } \frac{\delta S}{\delta b^i_{\ \mu}} \varepsilon^i_{\ jk} b^j_{\ \mu} + \frac{\delta S}{\delta \omega^i_{\ \mu}} \varepsilon^i_{\ jk} \omega^j_{\ \mu} -\partial_\mu\left(\frac{\delta S}{\delta \omega^k_{\ \mu}}\right) = 0\\
&{\ } \frac{\delta S}{\delta b^i_{\ \mu}} \partial_\lambda b^i_{\ \mu} + \frac{\delta S}{\delta \omega^i_{\ \mu}} \partial_\lambda \omega^i_{\ \mu} - \partial_\mu \left(b^i_{\ \lambda}\frac{\delta S}{\delta b^i_{\ \mu}} + \omega^i_{\ \lambda} \frac{\delta S}{\delta \omega^i_{\ \mu}} \right)=0\\
\end{split}
\end{eqnarray}
If these identities can be mapped to the identities following from the gauge transformations (\ref{field transf gauge}) one would say that the symmetries are equivalent. Otherwise they are inequivalent.

From the invariance of (\ref{action}) under (\ref{field transf gauge}), we find,
\begin{eqnarray}
\label{Gauge gen taylor}
&{\!}& S\left[b^i_{\ \mu} , \omega^i_{\ \mu}\right] = S\left[b^i_{\ \mu} + \delta_{\scriptscriptstyle\text{Gauge}}b^i_{\ \mu} , \omega^i_{\ \mu} + \delta_{\scriptscriptstyle\text{Gauge}} \omega^i_{\ \mu}\right]\nonumber\\
\Rightarrow &{\,}& S\left[b^i_{\ \mu} ,\omega^i_{\ \mu}\right] = S\left[b^i_{\ \mu} + \left(\partial_\mu \epsilon^i + \varepsilon^i_{\ jk}\omega^j_{\ \mu} \epsilon^k - p\,\varepsilon^i_{\ jk}b^j_{\ \mu} \epsilon^k + \varepsilon^i_{\ jk}b^j_{\ \mu} \tau^k \right),\right.\nonumber\\
&{\!}&\left. \qquad\qquad\qquad\quad\, \omega^i_{\ \mu} +\left(\partial_\mu \tau^i + \varepsilon^i_{\ jk}\omega^j_{\ \mu} \tau^k - q \varepsilon^i_{\ jk}b^j_{\ \mu} \epsilon^k \right) \right]\nonumber\\
&{\!}& \qquad \qquad\quad\,  = S\left[b^i_{\ \mu} , \omega^i_{\ \mu}\right] - \int d^3x ~\frac{\delta S}{\delta b^i_{\ \mu}} ~\left(\partial_\mu \epsilon^i + \varepsilon^i_{\ jk}\omega^j_{\ \mu} \epsilon^k - p\varepsilon^i_{\ jk}b^j_{\ \mu} \epsilon^k + \varepsilon^i_{\ jk}b^j_{\ \mu} \tau^k \right)\nonumber\\
&{\!}& \qquad \qquad \qquad \ - \int d^3x ~\frac{\delta S}{\delta \omega^i_{\ \mu}} ~\left(\partial_\mu \tau^i + \varepsilon^i_{\ jk}\omega^j_{\ \mu} \tau^k - q \varepsilon^i_{\ jk}b^j_{\ \mu} \epsilon^k \right)\nonumber\\
\Rightarrow &{\!}&  \int d^3x \left[-\partial_\mu\left(\frac{\delta S}{\delta \omega^k_{\ \mu}}\right) + \frac{\delta S}{\delta b^i_{\ \mu}} \varepsilon^i_{\ jk} b^j_{\ \mu} + \frac{\delta S}{\delta b^i_{\ \mu}} \varepsilon^i_{\ jk} \omega^j_{\ \mu} \right] \tau^k \nonumber\\
&{\!}& \qquad + \int d^3x \left[-\partial_\mu\left(\frac{\delta S}{\delta b^k_{\ \mu}}\right) - q ~\frac{\delta S}{\delta \omega^i_{\ \mu}} \varepsilon^i_{\ jk} b^j_{\ \mu} + \frac{\delta S}{\delta b^i_{\ \mu}} \varepsilon^i_{\ jk} \omega^j_{\ \mu} -p \frac{\delta S}{\delta b^i_{\ \mu}} \varepsilon^i_{\ jk} b^j_{\ \mu} \right] \epsilon^k = 0
\end{eqnarray}
which leads to the independent gauge identities:
\begin{eqnarray}
\label{gauge ident generator}
\begin{split}
&{\ } -\partial_\mu\left(\frac{\delta S}{\delta \omega^k_{\ \mu}}\right) + \frac{\delta S}{\delta b^i_{\ \mu}} \varepsilon^i_{\ jk} b^j_{\ \mu} + \frac{\delta S}{\delta b^i_{\ \mu}} \varepsilon^i_{\ jk} \omega^j_{\ \mu}=0\\
&{\ } -\partial_\mu\left(\frac{\delta S}{\delta b^k_{\ \mu}}\right) - q ~\frac{\delta S}{\delta \omega^i_{\ \mu}} \varepsilon^i_{\ jk} b^j_{\ \mu} + \frac{\delta S}{\delta b^i_{\ \mu}} \varepsilon^i_{\ jk} \omega^j_{\ \mu} -p \frac{\delta S}{\delta b^i_{\ \mu}} \varepsilon^i_{\ jk} b^j_{\ \mu}=0.\\
\end{split}
\end{eqnarray}
Comparing (\ref{gauge ident generator}) with (\ref{gauge ident PGT}) it is easy to be convinced that they are inequivalent. Also note that all the identities become trivial as we invoke the equations of motion.

In the last paragraph, we have shown the {\it{inequivalence}} between the Poincare gauge invariance and the gauge invariance by using the lagrangian identities. One is then naturally led to the question as to what happens to the 2nd order metric gravity, where the equivalence between the spacetime and gauge invariances was demonstrated canonically with an off-shell map (\ref{Munattended} and \ref{M}). To further elucidate the question of symmetry it will, therefore, be useful to reconsider the symmetries of the 2nd order metric gravity from the point of view of such identities. For convenience we will start from (\ref{L}) which is equivalent to the Einstein action (\ref{Einsteinaction}) as mentioned earlier. The basic fields are $N^{\bot}$, $N^\alpha$ and $g_{\alpha\beta}$. Their gauge variations are given by equations (\ref{GV1}), (\ref{GV2}) and (\ref{gGV}) respectively. The identities following from the symmetry of (\ref{L}) under these transformations are:
\begin{eqnarray}
W^{\scriptscriptstyle{\text{gauge}}}=-\frac{d}{dt}\left(\frac{\delta S}{\delta N^{\bot}}\right) + \partial_\gamma\left(N^\gamma\frac{\delta S}{\delta N^{\bot}}\right) + \partial_\gamma\left(g^{\gamma\alpha}N^{\bot}\frac{\delta S}{\delta N^{\alpha}}\right)&{\ }& \label{GI1} \nonumber \\
+\ g^{\gamma\alpha}\partial_\gamma N^{\bot}\frac{\delta S}{\delta N^{\alpha}} \!\! &-& \!\! 2K_{\alpha\beta}\frac{\delta S}{\delta g_{\alpha\beta}} = 0,
\end{eqnarray}
\begin{eqnarray}
W_\alpha^{\scriptscriptstyle{\text{gauge}}}=-\frac{d}{dt}\left(\frac{\delta S}{\delta N^{\alpha}}\right) + \partial_\gamma\left(N^\gamma\frac{\delta S}{\delta N^{\alpha}}\right) + \partial_\alpha N^{\bot}\frac{\delta S}{\delta N^{\bot}} \!\! &+& \!\! \partial_\alpha N^{\gamma}\frac{\delta S}{\delta N^{\gamma}} \label{GI2} \nonumber \\
+\,\  \partial_\alpha g_{\gamma\beta}\frac{\delta S}{\delta g_{\gamma\beta}}  \!\! &-& \!\! 2\partial_\beta\left(g_{\alpha\gamma}\frac{\delta S}{\delta g_{\gamma\beta}}\right)=0.
\end{eqnarray}
Likewise, the identities corresponding to the diff invariances (\ref{R1}), (\ref{R2}) and (\ref{gR}) can similarly be worked out. The identity corresponding to $\xi^0$ is
\begin{eqnarray}
W^{\scriptscriptstyle{\text{diff}}}=-\frac{d}{dt}\left(\frac{\delta S}{\delta N^{\bot}}\right)N^{\bot} + N^{\bot}\partial_\gamma\left(N^\gamma\frac{\delta S}{\delta N^{\bot}}\right) + N^{\gamma}\partial_\gamma N^{\bot}\frac{\delta S}{\delta N^{\bot}} - \frac{d}{dt}\left(\frac{\delta S}{\delta N^{\alpha}}\right)N^\alpha \label{RI1} \nonumber \\
+ N^{\alpha}\partial_\gamma\left(N^\gamma\frac{\delta S}{\delta N^{\alpha}}\right) + N^\gamma\partial_\gamma N^\alpha\frac{\delta S}{\delta N^{\alpha}}+ \partial_\beta\left(\frac{\delta S}{\delta N^{\alpha}} (N^{\bot})^2 g^{\alpha\beta}\right)\nonumber \\
+ \frac{\delta S}{\delta g_{\alpha\beta}}{\dot{g_{\alpha\beta}}} - 2\partial_\beta\left(N_\alpha\frac{\delta S}{\delta g_{\alpha\beta}}\right) = 0,
\end{eqnarray}
and those corresponding to $\xi^\alpha$ are
\begin{eqnarray}
W_\alpha^{\scriptscriptstyle{\text{diff}}}=-\frac{d}{dt}\left(\frac{\delta S}{\delta N^{\alpha}}\right) + \partial_\gamma\left(N^\gamma\frac{\delta S}{\delta N^{\alpha}}\right) + \partial_\alpha N^{\bot}\frac{\delta S}{\delta N^{\bot}} + \partial_\alpha N^{\gamma}\frac{\delta S}{\delta N^{\gamma}} + \partial_\alpha g_{\gamma\beta}\frac{\delta S}{\delta g_{\gamma\beta}} \label{RI2} \nonumber \\
-2\partial_\beta\left(g_{\alpha\gamma}\frac{\delta S}{\delta g_{\gamma\beta}}\right) =0.
\end{eqnarray}
The identities (\ref{RI2}) are identical with (\ref{GI2}) (i.e. $W_\alpha^{\scriptscriptstyle{\text{diff}}} = W_\alpha^{\scriptscriptstyle{\text{gauge}}}$) while (\ref{RI1}) are apparently different from (\ref{GI1}). However, a little algebra shows
\begin{eqnarray*}
N^\perp W^{\scriptscriptstyle{\text{gauge}}} + N^\alpha W_\alpha^{\scriptscriptstyle{\text{gauge}}} = W^{\scriptscriptstyle{\text{diff}}}.
\end{eqnarray*}
The set of gauge identities (\ref{GI1}) and (\ref{GI2}) is thus equivalent to the set (\ref{RI1}) and (\ref{RI2}) following from reparametrization invariances. This is consistent with our canonical analysis of 2nd order metric gravity where we demonstrated the equivalence between the gauge and diff parameters by devising the one to one mapping (\ref{Munattended}, \ref{M}).

\section{Conclusion}

Recently the 3D gravity models in the framework of Poincare gauge theory (PGT) have come to forefront \cite{Blagojevic:2009ek, Blagojevic:2008bn, Blagojevic:2004hj, Blagojevic:2003uc, Blagojevic:2002du, Carlip:2005zn, Park:2008yy, Grumiller:2008pr, Carlip:2008qh} in the literature. Among the various issues considered, a particularly significant one is the difference between the PGT transformations of the basic fields and the gauge variations of the same obtained in the canonical way. The two can only be mapped using the equations of motion. This fact was observed earlier \cite{ Blagojevic:2004hj, Blagojevic:2003uc} but its significance was missed, principally due to the fact that the canonical gauge generator was constructed following \cite{Castellani:1981us} which maps solutions to solutions of the equations of motion. We have shown here, in the context of the topological 3D gravity with torsion, that the general gauge transformations can be obtained in the canonical way in an off-shell manner. This is done by following a method available in the literature \cite{Banerjee:1999hu, Banerjee:1999yc} that views the gauge transformations as mapping field configurations to field configurations. This naturally lends a new perspective to this issue of symmetries.

The PGT formalism is reviewed and the geometric interpretation is  scrutinised by establishing the basic PGT transformations geometrically, using general coordinate (diff) transformations and local Lorentz transformations. The off-shell invariance of the model under PGT transformations has been explicitly verified. Then a complete canonical analysis of the model is presented. This model presents an example of a mixed constrained system with both first and second class constraints. The reduced phase space is obtained by completely eliminating the second class constraints using Dirac brackets. Use of Lagrange multipliers, as done in \cite{Blagojevic:2004hj, Blagojevic:2003uc}, is thereby avoided. The model then transpires to a standard gauge system having only first class constraints. The difference is that the symplectic structure is defined by the Dirac brackets instead of the usual Poisson brackets. The generator of gauge transformations that map field configurations to field configurations, is constructed by following the structured algorithm given in \cite{Banerjee:1999hu, Banerjee:1999yc}. We find the transformations of the basic fields by computing their Dirac brackets with the gauge generator and check by direct calculation that these gauge transformations are again  {\it{off-shell}} invariances of the action. The gauge transformations of the basic fields are then compared with the analogous transformations under PGT. There exists no off-shell map between them, though the two agree on-shell.

To put our findings in a proper perspective, we carry out a similar analysis for 2+1 dimensional Einstein gravity in the usual metric formulation. In this case we prove an exact off-shell equivalence of the general coordinate (diff) transformations with the gauge transformations found by a canonical (hamiltonian) approach. This clearly manifests the peculiarity of the PGT vis-a-vis a standard gauge theory.

Finally, a lagrangian analysis of symmetries based on identities was performed. Whereas in the hamiltonian treatment one has to find a map between the parameters, in the lagrangian analysis there should be a map that connects the identities which involve the basic variables of the theory. In the framework of PGT, it was shown that the identities were different for the two types of symmetries. A mapping of the identities was not possible thereby reconfirming the results from the hamiltonian formalism. For the Einstein gravity, on the contrary, a mapping between the identities was explicitly derived. Our analysis shows that the hamiltonian and lagrangian formulations actually complement one another. For discussing off-shell equivalence, the lagrangian approach is more practical since it becomes obvious that a map between the identities cannot exist. In the hamiltonian formulation, it is nontrivial to really prove that a map does not exist between the transformation parameters. For discussing on-shell equivalence, on the contrary, the lagrangian method is not appropriate since the identities trivialise (0=0). Here the hamiltonian approach is clearly more viable. As a final remark, we mention that the methods developed here may be applied to other 3D gravity models \cite{Blagojevic:2009ek, Blagojevic:2008bn, Blagojevic:2004hj, Blagojevic:2003uc, Carlip:2005zn, Park:2008yy, Grumiller:2008pr, Carlip:2008qh}.


\begin{appendix}

\renewcommand{\thesection}{Appendix \Alph{section}}			
\setcounter{section}{0}										

\section{:\ \ The Poisson algebra of constraints}
\label{App:Poisson}

\renewcommand{\theequation}{A.\arabic{equation}} 
\setcounter{equation}{0}  

The basic non-zero Poisson brackets of the theory (\ref{action}) are given below.
\begin{eqnarray}
\label{fieldPoisson}
\begin{split}
\lbrace b^i_{\ \mu} (x), \pi^{\ \nu}_j (x') \rbrace &=& \delta^i_j ~\delta^\nu_\mu ~\delta(x-x')\\
\lbrace \omega^i_{\ \mu} (x), \Pi^{\ \nu}_j (x') \rbrace &=& \delta^i_j ~\delta^\nu_\mu ~\delta(x-x')\\
\end{split}
\end{eqnarray}
Also, we give below a list of the Poisson brackets of the quantities $\mathcal{H}$ and $\mathcal{K}$, constructed out of the the basic fields in (\ref{canon Hamilt}), with the primary constraints.
\begin{eqnarray}
\begin{split}
\label{App:SRel PAlgebra}
\lbrace\phi_i^{\ \alpha}(x),\mathcal{H}_j(x')\rbrace &= 2 \,\varepsilon^{0\alpha\beta}\left[\alpha_4 \,\eta_{ij}\, \partial^{(x)}_\beta \delta(x-x')-\varepsilon_{ijk}\left(\alpha_4\,\omega^k_{\ \beta}-\Lambda \,b^k_{\ \beta} \right)\delta(x-x') \right]\\
\lbrace\phi_i^{\ \alpha}(x),\mathcal{K}_j(x')\rbrace &= 2 \,\varepsilon^{0\alpha\beta}\left[a \,\eta_{ij} \,\partial^{(x)}_\beta \delta(x-x')-\varepsilon_{ijk}\left(a\,\omega^k_{\ \beta} + \alpha_4 \,b^k_{\ \beta} \right)\delta(x-x') \right]\\
\lbrace\Phi_i^{\ \alpha}(x),\mathcal{H}_j(x')\rbrace &= 2 \,\varepsilon^{0\alpha\beta}\left[a \,\eta_{ij} \,\partial^{(x)}_\beta \delta(x-x')-\varepsilon_{ijk}\left(a\,\omega^k_{\ \beta} + \alpha_4 \,b^k_{\ \beta} \right)\delta(x-x') \right]\\
\lbrace\phi_i^{\ \alpha}(x),\mathcal{K}_j(x')\rbrace &= 2 \,\varepsilon^{0\alpha\beta}\left[\alpha_3 \,\eta_{ij} \,\partial^{(x)}_\beta \delta(x-x')-\varepsilon_{ijk}\left(\alpha_3\,\omega^k_{\ \beta} + a \,b^k_{\ \beta} \right)\delta(x-x') \right].
\end{split}
\end{eqnarray}

We now calculate the non-trivial Poisson algebra of the constraints, by using the algebra among basic variables  (\ref{fieldPoisson}). The algebra (\ref{App:SRel PAlgebra}) comes in handy at this step (as well as in the following calculations).
\begin{eqnarray}
\label{App:PP PAlgebra}
\begin{split}
\lbrace\phi_i^{\ \alpha}(x),\phi_j^{\ \beta}(x')\rbrace &= -2 \,\alpha_4 \,\varepsilon^{0\alpha\beta}\,\eta_{ij}\,\delta(x-x')\\
\lbrace\Phi_i^{\ \alpha}(x),\Phi_j^{\ \beta}(x')\rbrace &= -2 \,\alpha_3 \,\varepsilon^{0\alpha\beta}\,\eta_{ij}\,\delta(x-x')\\
\lbrace\phi_i^{\ \alpha}(x),\Phi_j^{\ \beta}(x')\rbrace &= -2 \,a \,\varepsilon^{0\alpha\beta}\,\eta_{ij}\,\delta(x-x')\\
\end{split}
\end{eqnarray}
Observe that the Poisson algebra (\ref{App:PP PAlgebra}) between the primary constraints does not close, implying the existence of second-class constraints.

The Poisson algebra between primary and secondary constraints are:
\begin{eqnarray}
\begin{split}
\label{App:PS PAlgebra}
\lbrace\phi_i^{\ \alpha}(x),\bar{\mathcal{H}_j}(x')\rbrace &= \varepsilon_{ijk} \left(p\,\phi^{k\alpha} + q\,\Phi^{k\alpha} \right) \delta(x-x')\\
\lbrace\phi_i^{\ \alpha}(x),\bar{\mathcal{K}_j}(x')\rbrace &= -\varepsilon_{ijk}\,\phi^{k\alpha} \,\delta(x-x')\\
\lbrace\Phi_i^{\ \alpha}(x),\bar{\mathcal{H}_j}(x')\rbrace &= -\varepsilon_{ijk}\,\phi^{k\alpha} \,\delta(x-x')\\
\lbrace\Phi_i^{\ \alpha}(x),\bar{\mathcal{K}_j}(x')\rbrace &= -\varepsilon_{ijk}\,\Phi^{k\alpha} \,\delta(x-x'),\\
\end{split}
\end{eqnarray}
while the algebra among the secondary constraints are:
\begin{eqnarray}
\begin{split}
\label{App:SS PAlgebra}
\lbrace\bar{\mathcal{H}_i}(x),\bar{\mathcal{H}_j}(x')\rbrace &= \varepsilon_{ijk}\left(p\,\bar{\mathcal{H}^k}+q\,\bar{\mathcal{K}^k}\right)\delta(x-x')\\
\lbrace\bar{\mathcal{K}_i}(x),\bar{\mathcal{K}_j}(x')\rbrace &= - \varepsilon_{ijk}\,\bar{\mathcal{K}^k}\,\delta(x-x')\\
\lbrace\bar{\mathcal{H}_i}(x),\bar{\mathcal{K}_j}(x')\rbrace &= - \varepsilon_{ijk}\,\bar{\mathcal{H}^k}\,\delta(x-x').\\
\end{split}
\end{eqnarray}
We see that both sets (\ref{App:PS PAlgebra}, \ref{App:SS PAlgebra}) close.

\section{:\ \ On the significance of (\ref{RB master 1})}
\label{App:2nd Master}

\renewcommand{\theequation}{B.\arabic{equation}} 	
\setcounter{equation}{0}  							

In this appendix, we would like to make a note on the information content of (\ref{RB master 1}), which is referred to as the ``first condition" hereafter. This equation gives the variation of the Lagrange multipliers corresponding to the primary (first-class) constraints in terms of the structure constants $\left(V^I_{\;\;\, J}\right)(x,x')$ and $\left(C^I_{\;\; JK}\right)(x,x',x'')$ defined in (\ref{StructureConsts}). However, as we show below, this equation gives us no new restrictions on the parameters. This is because the equation itself can be obtained from the properties of the total hamiltonian and the second condition -- the `master equation' (\ref{RB master 2}) \cite{Banerjee:1999hu}. We now demonstrate this fact here, in the context of our theory.

We begin by calculating the time variation of the field $\displaystyle b^i_{\ 0}$ by taking its Dirac bracket\footnote{Recall that we have adopted the approach of eliminating all second-class constraints by using Dirac brackets. Hence equations of motion are given by Dirac brackets.} with the total hamiltonian (\ref{total Hamilt}), to see that it gives the Lagrange multiplier corresponding to $\displaystyle \pi^{\ 0}_i$,
\begin{eqnarray}
\label{b lambda 3}
\dot{b}^i_{\ 0} = \lbrace b^i_{\ 0}, \int d^2x \,\mathcal{H}_T \rbrace^* = \lambda^{(3)i}_{\;\quad 0}.
\end{eqnarray}
Using this, we find the variation of the multiplier $\lambda^{(3)}$ in terms of the derivative of the field transformations,
\begin{eqnarray}
\label{field lambda 3}
\delta \lambda^{(3)i}_{\;\quad 0} = \delta \dot{b}^i_{\ 0} = \frac{d}{dt} \delta b^i_{\ 0}.
\end{eqnarray}
However, we have already calculated the transformation $\delta b^i_{\ 0}$ (\ref{field transf gauge}). Also, recall that \emph{only} the `master equation' (\ref{RB master 2}) was required in deriving the generator, and so, (\ref{field transf gauge}) is independent of the first condition. Now substituting these field transformations for $b^i_{\ 0}(x)$ in the last equation (\ref{field lambda 3}), and using the definitions $\epsilon=\epsilon^{(1)}$ and $\tau=\epsilon^{(2)}$ introduced before in Section 5, we get:
\begin{eqnarray}
\label{intermediate 3}
\frac{d}{dt} \delta b^i_{\ 0} = \frac{d}{dt} \left[\partial_0 \epsilon^{(1)i} + \varepsilon^i_{\ jk}\,\omega^j_{\ 0} \epsilon^{(1)k}-p\,\varepsilon^i_{\ jk}\,b^j_{\ 0}\epsilon^{(1)k} + \varepsilon^i_{\ jk}\,b^j_{\ 0}\epsilon^{(2)k}\right].
\end{eqnarray}
This can be related with the variation of $\lambda^{(3)}$ by taking advantage of (\ref{field lambda 3}), to finally obtain
\begin{eqnarray}
\label{from theory 3}
\delta \lambda^{(3)i}_{\;\quad 0} = \frac{d}{dt} \left[ \dot{\epsilon}^{(1)i} - \epsilon^{(1)k}\,\varepsilon_k^{\ \,ij} \left(p\,b_{j0} - \omega_{j0}\right) + \epsilon^{(2)k}\varepsilon_k^{\ \,ij}\,b_{j0} \right] = \frac{d}{dt} \epsilon^{(3)i}.
\end{eqnarray}
Here, in the last step, we have used (\ref{rel epsilons}) to express $\epsilon^{(1)}$ and $\epsilon^{(2)}$ in terms of $\epsilon^{(3)}$.

Let us now return to the first condition (\ref{RB master 1}), from which it follows,
\begin{eqnarray}
\label{1st master 3}
\delta \lambda^{(3)}(x) &=& \displaystyle\frac{d\epsilon^{(3)}(x)}{dt}-\int d^2x' \,\epsilon^{(I)}(x') \,\left[ \left(V^3_{\;\;\: I}\right)(x,x') +\int d^2x''\,\lambda^{(B)}(x'') \,\left(C^3_{\;\; IB}\right)(x,x',x'')\right]\nonumber\\
&=& \frac{d\epsilon^{(3)}(x)}{dt}
\end{eqnarray}
thereby reproducing (\ref{from theory 3}). The second term does not contribute since the structure constants $\left(V^3_{\;\;\: I}\right)$, $\left(C^3_{\;\; IB}\right)$ vanish (\ref{Cs} \& \ref{Vs}). This shows that, as claimed at the beginning of this appendix, the first condition gives us no new restrictions on the parameters. It is basically a consequence of (\ref{RB master 2}).

In our calculations above, we have used the Lagrange multiplier $\lambda^{(3)}$ corresponding to $\pi^{\ 0}_i$. However, by the same process, analogous results are obtained for the multiplier $\lambda^{(4)}$ which corresponds to $\Pi^{\ 0}_i$. The starting point of the calculation for $\lambda^{(4)}$ is now:
\begin{eqnarray}
\dot{\omega}^i_{\ 0} = \lbrace \omega^i_{\ 0}, \int d^2x \,\mathcal{H}_T \rbrace^* = \lambda^{(4)i}_{\quad\; 0}.
\end{eqnarray}
Then, going through similar steps analogous to (\ref{field lambda 3}, \ref{intermediate 3} \& \ref{from theory 3}) we get
\begin{eqnarray}
\label{from theory 4}
\delta \lambda^{(4)i}_{\quad\; 0} = \frac{d}{dt} \left[ \dot{\epsilon}^{(2)i} - q \,\epsilon^{(1)k}\,\varepsilon_k^{\ ij}\, b_{j0} +  \epsilon^{(2)k}\,\varepsilon_k^{\ ij}\,\omega_{j0} \right] = \frac{d}{dt} \epsilon^{(4)i}.
\end{eqnarray}
which is the analogue of (\ref{from theory 3}) found above. This is nothing but the first condition (\ref{RB master 1}) corresponding to  $\lambda^{(4)}$,
\begin{eqnarray}
\label{1st master 4}
\delta \lambda^{(4)}(x) &=& \displaystyle\frac{d\epsilon^{(4)}(x)}{dt}-\int d^2x' \,\epsilon^{(I)}(x') \,\left[ \left(V^4_{\;\;\: I}\right)(x,x') +\int d^2x''\,\lambda^{(B)}(x'') \,\left(C^4_{\;\; IB}\right)(x,x',x'')\right] \nonumber\\
&=& \frac{d\epsilon^{(4)}(x)}{dt},
\end{eqnarray}
which follows as a consequence of the vanishing of the structure constants $\left(V^4_{\;\;\, J}\right)$ and $\left(C^4_{\;\; JK}\right)$ calculated in (\ref{Cs} \& \ref{Vs}).

\end{appendix}

\section*{Acknowledgement}

Two of the authors, SG and PM, would like to acknowledge the facilities extended to them during their visit to the S.N.Bose National Centre for Basic Sciences, Kolkata as visiting associates. PM also likes to acknowledge University Grants Commission  for support through the project no. F.PSW-027/07-08 (ERO).


\end{document}